\documentclass[12pt]{iopart}
\usepackage{graphicx}
\usepackage{subcaption}
\usepackage{xcolor}
\usepackage[hidelinks]{hyperref}
\usepackage{siunitx}
\usepackage{comment}
\usepackage{cite}
\usepackage{lineno}

\hypersetup{
    colorlinks=true,
    linkcolor=blue,
    filecolor=blue,      
    urlcolor=blue,
    citecolor=blue,
}
\definecolor{darkgreen}{rgb}{0.0, 0.4, 0.0}

\begin{document}

\title{Using t-SNE for characterizing glitches in LIGO detectors}

\author{Tabata Aira Ferreira, Gabriela González}

\address{Louisiana State University, Baton Rouge, LA 70803, USA}
\ead{tferreira@lsu.edu}
\vspace{10pt}
\begin{indented}
\item[]
\end{indented}

\begin{abstract} Glitches are non-Gaussian noise transients originating from environmental and instrumental sources that contaminate data from gravitational wave detectors. Some glitches can even mimic gravitational wave signals from compact object mergers, which are the primary targets of terrestrial observatories. In this study, we present a method to analyze noise transients from the LIGO observatories using $Q$-transform information combined with t-Distributed Stochastic Neighbor Embedding (t-SNE). We implement classification techniques, examine the influence of parameters on glitch classification, and conduct a week-long daily analysis to track outlier transients over time.

\end{abstract}


\section{Introduction}

A new era of gravitational-wave astronomy began on September 14, 2015, with the first detection of a binary black hole merger~\cite{abbott2016observation}. The latest published catalog, covering the first three LIGO-Virgo-KAGRA observing runs between 2015 and 2020~\cite{abbott2023gwtc}, included 90 detections. The LVK collaboration - composed of the LIGO (Laser Interferometer Gravitational-Wave Observatory)\cite{aasi2015advanced,Harry:2010zz,martynov2016sensitivity}, Virgo\cite{acernese2014advanced}, and KAGRA~\cite{akutsu2021overview,akutsu2019kagra,aso2013interferometer} teams - is actively engaged in the search for new detections and a deeper understanding of the Universe. Currently, the observatories are in their fourth observing run (O4), which began on May 24, 2023, and is divided into three parts. O4a (ending on January 16, 2024), which is the focus of this work, included 81 public alerts corresponding to confident detection candidates~\cite{graceDBo4a}. O4b (from April 10, 2024, to January 28, 2025) resulted in 105 additional confident alerts~\cite{graceDBo4b} and was followed by O4c, which is ongoing at the time of writing.

The observatories are constantly being upgraded to achieve the best possible sensitivity through experimental tests and instrumental changes. However, with increased sensitivity, they also become more susceptible to \textit{glitches}, non-Gaussian noise transients that have the potential to contaminate gravitational wave data. As the detectors become more sensitive, non-astrophysical transients with the same amplitude can produce signals with a higher signal-to-noise ratio, emphasizing the need for effective glitch identification and mitigation strategies.

Glitches are usually associated with environmental or instrumental sources; during the third observing run (O3), for instance, LIGO was significantly affected by glitches from light scattering~\cite{soni2020reducing,soni2024modeling}, which contaminated the data mainly between 20 and \SI{40}{\Hz} during high-ground motion periods. 

In general, these noise transients pollute the data, affecting the search for gravitational waves, increasing the background noise in the statistical analysis, and making difficult the search for \textit{burst} signals, which are short in duration, unmodeled, and are expected to originate from events such as core-collapse supernovae~\cite{abbott2021all}. Some glitches may resemble real signals, potentially leading to false positives, while others may coincide with actual gravitational wave signals, as occurred during the first detection of a binary neutron star merger, GW170817~\cite{abbott2017gw170817}.

Glitches are typically visualized via spectrograms, from which we can derive characteristics such as frequencies, duration, morphology, and signal-to-noise ratio (SNR). The study of glitches is one of the responsibilities of the Detector Characterization (DetChar) group. To identify glitch sources, DetChar frequently relies on auxiliary channels, which record data from sensors distributed throughout the interferometers~\cite{davis2021ligo,acernese2023virgo}. These sensors include photodiodes, magnetometers, seismometers, thermometers, microphones, and others. Temporal (and sometimes morphological) coincidences between these sensors and the gravitational wave channel, where LIGO stores information about potential gravitational wave candidates, can help indicate the source of glitches. Additionally, the lack of such coincidences assist in validating gravitational wave events~\cite{davis2021ligo,soni2024ligo}.

Understanding the behavior of glitches is crucial for identifying their sources and hopefully eliminating or reducing their occurrences. LIGO looks for glitches by using {\tt Omicron}, a tool that performs a multi-resolution analysis of gravitational data in the time-frequency domain~\cite{robinet2020omicron,GWOLLUM}. Omicron processes the data via the $Q$-transform~\cite{brown1991calculation} in two stages. First, it analyzes transients across different time-frequency planes (also known as $Q$-planes), which vary according to a quality factor parameter, $Q$. This information is typically stored in what we will refer to as ``unclustered files''. Subsequently, after clustering over time, Omicron stores information for each glitch with representative parameters such as time, frequency, duration, SNR, and others. From clustered Omicron information, tools like {\tt Gravity Spy}~\cite{zevin2017gravity,glanzer2023data,zevin2024gravity} are able to classify glitches by analyzing their morphologies in spectrograms.

Several investigations have been conducted to analyze and classify glitches in gravitational wave data. In~\cite{george2018classification}, for instance, the authors present the first application of deep learning combined with transfer learning to classify glitches. Their method reduces training time and employs t-SNE for unsupervised clustering, helping identify new glitch classes. \texttt{GWSkyNet}, on the other hand, is a tool designed to distinguish in real time between glitches and astrophysical events~\cite{cabero2020gwskynet,chan2024gwskynet}. Other approaches have applied Principal Component Analysis (PCA) or machine learning techniques for noise classification, with performance demonstrated using simulated transients~\cite{powell2017classification}. There is also a tool, \texttt{GSpyNetTree}, which employs a Convolutional Neural Network to classify astrophysical signals versus glitches in gravitational wave event candidates~\cite{alvarez2024gspynettree}. All these efforts aim to advance the characterization and classification of glitches.

This paper presents a method for characterizing glitches using t-SNE~\cite{van2008visualizing, van2009learning}, t-Distributed Stochastic Neighbor Embedding, on unclustered Omicron data. We provide examples of the method's application in LIGO during O3b, the second half of the third observing run, and explore its use during O4a and O4c. Since Omicron already runs daily, there is no need to generate new spectrograms or new data for this analysis. Additionally, this method does not require the creation of a training dataset and can analyze glitches with low latency (within a few minutes) as it relies solely on Omicron.

In the context of glitch analysis, this unsupervised technique has been applied in various ways, including visualizing the Gravity Spy dataset using lower-resolution spectrograms~\cite{bahaadini2018machine}, examining compression through autoencoders for anomaly detection in LIGO's glitch populations~\cite{laguarta2024detection}, identifying unknown classes of glitches and anomalous signals via transfer learning~\cite{george2018classification}, and comparing results before and after modifications, such as improving interpretability using weights derived from attention modules, which were proposed and implemented as part of Gravity Spy during O4 \cite{wu2024advancing}. These applications, combined with the proven effectiveness of t-SNE in visualizing glitch groups from Omicron data~\cite{ferreira2022comparison}, motivated us to further explore the technique in different contexts and with distinct goals.

Section~\ref{sec:transient_noise} provides an overview of transient noise features, examines how Omicron unclustered data vary with $Q$-values, outlines the main classes of glitches, and introduces the method using the t-SNE algorithm. Section~\ref{sec:method_with_classes} discusses the O3b LIGO Livingston application and explores how parameter selection affects the results, using Gravity Spy classifications as a reference. Section~\ref{sec:random_data} presents results obtained without prior knowledge of glitch classes. The same approach is shown in Section~\ref{sec:one_week_data}, where outliers during O4a (at LLO) and O4c (at LHO) are identified and tracked over time, also without requiring prior knowledge of class label. Finally, Section~\ref{sec:conclusions} summarizes the conclusions and discusses perspectives for future work.

\section{Transient Noise}
\label{sec:transient_noise}

Transient noise, or glitches, create an excess background for astrophysical signals. To identify the presence of these unwanted signals, LIGO uses a tool called Omicron, which detects excess power in the main gravitational wave data channel. To identify these power excesses, Omicron applies a $Q$-transform to characterize these glitches based on their frequency, time, SNR, and $Q$-value, $Q$. The $Q$-value is related to the bisquare window in the $Q$-transform and proportional to the frequency window $f_0$, and its bandwidth $\Delta f_0$~\cite{chatterji2005search},

\begin{equation}
    Q \propto \frac{f_0}{\Delta f_0}.
    \label{eq:q_value}
\end{equation}

\begin{figure}[ht!]
    \centering
    \includegraphics[width=0.8\textwidth]{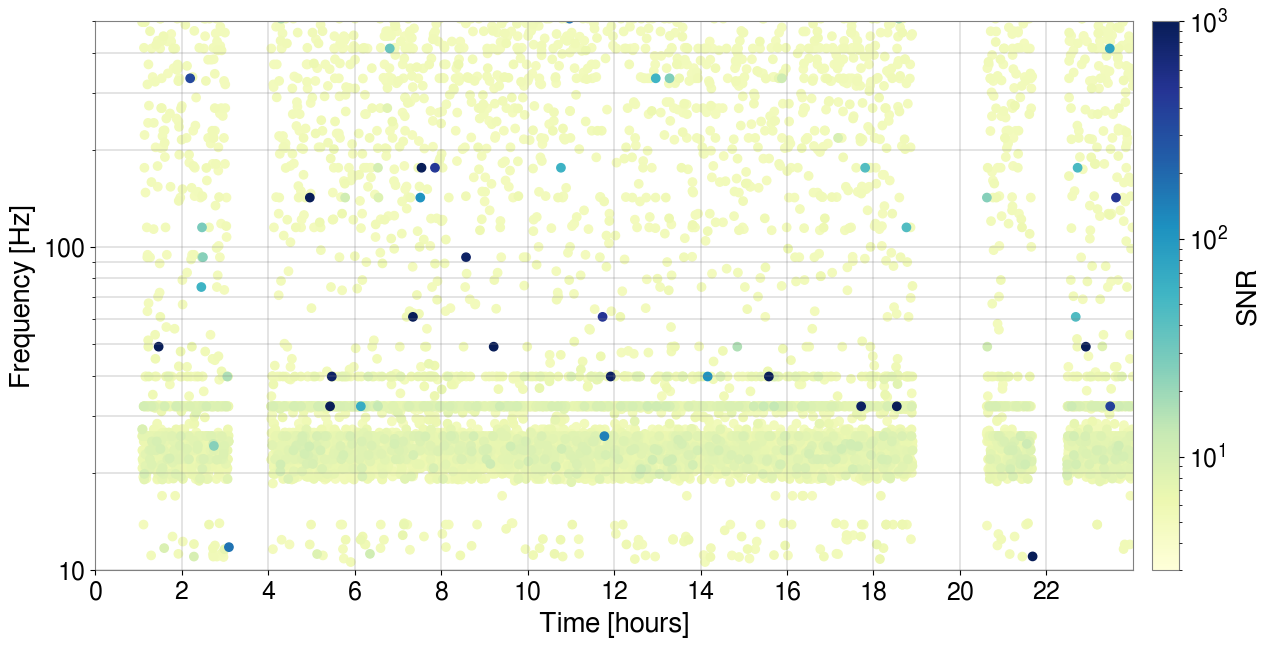}
    \caption{A one-day glitchgram during O4a at LHO (LIGO Hanford Observatory), built from Omicron clustered files. Each glitch is represented by a data point showing its frequency, time of occurrence, and SNR.}
    \label{fig:omicron_clustered_day}
\end{figure}

The primary advantage of using the $Q$-transform lies in its ability to vary the resolution in both time and frequency within a time-frequency representation. In the initial stage, Omicron saves the transient information in unclustered files (in ROOT format)~\cite{robinet2016omicron}, which includes storing parameters for all \textit{tiles} across various $Q$-values. In the context of spectrogram generation via the $Q$-transform, a \textit{tile} is defined as a rectangle with a height of ${\Delta f_0}$ and a width of ${\Delta t}$, built for different $Q$-values. The second step involves time clustering, where a single representative data point is selected for each transient. \Fref{fig:omicron_clustered_day} presents a \textit{glitchgram}, which is a time and frequency representation of glitches from clustered files over the period of a day. Information stored in the glitchgrams is crucial for understanding glitch behavior across different periods. On the day in~\Fref{fig:omicron_clustered_day}, there are significant broadband glitches between $20$ and \SI{40}{\Hz} with low SNR, as indicated by the color of the data points.

\begin{figure}[ht!]
    \centering
    \begin{subfigure}{0.32\textwidth}
        \centering
        \caption{$Q$-value = 4.56}
        \includegraphics[width=\textwidth]{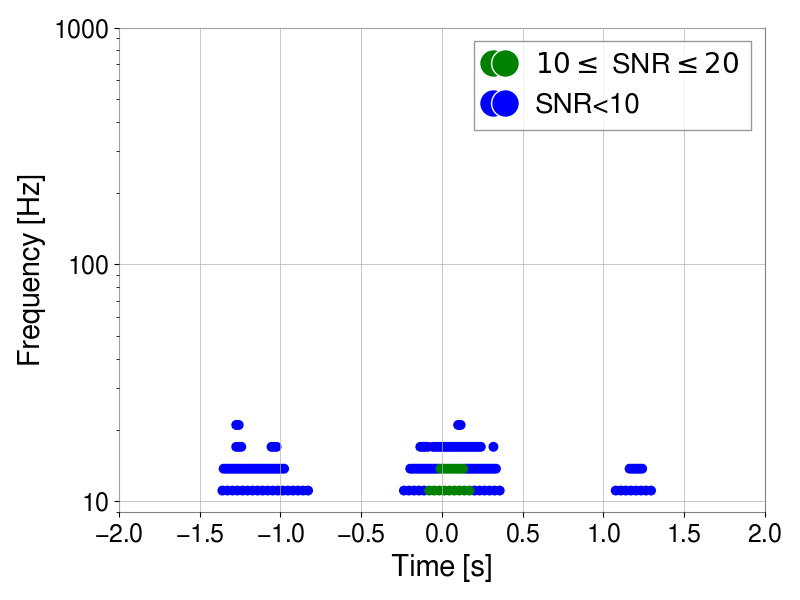}
        \label{fig:q4_56}
    \end{subfigure}
    \begin{subfigure}{0.32\textwidth}
        \centering
        \caption{$Q$-value = 8.60}
        \includegraphics[width=\textwidth]{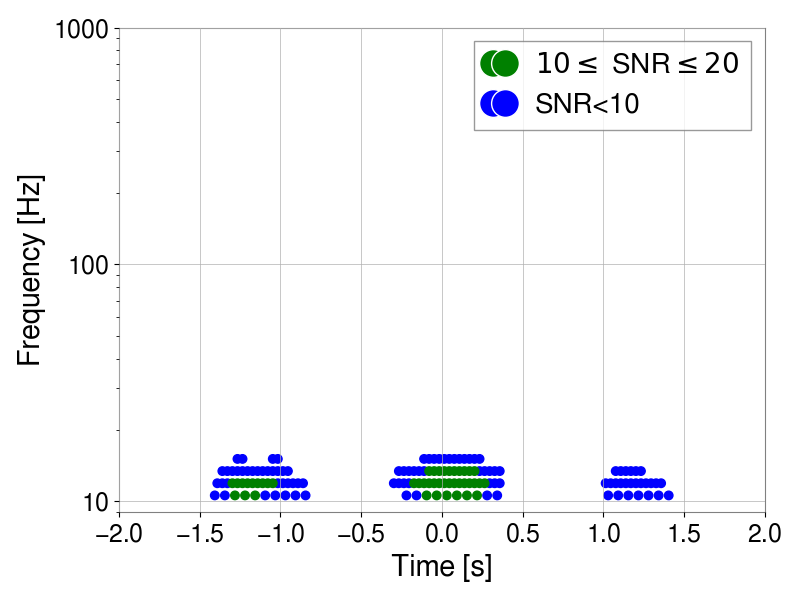}
        \label{fig:q8_60}
    \end{subfigure}
    \begin{subfigure}{0.32\textwidth}
        \centering
        \caption{$Q$-value = 16.23}
        \includegraphics[width=\textwidth]{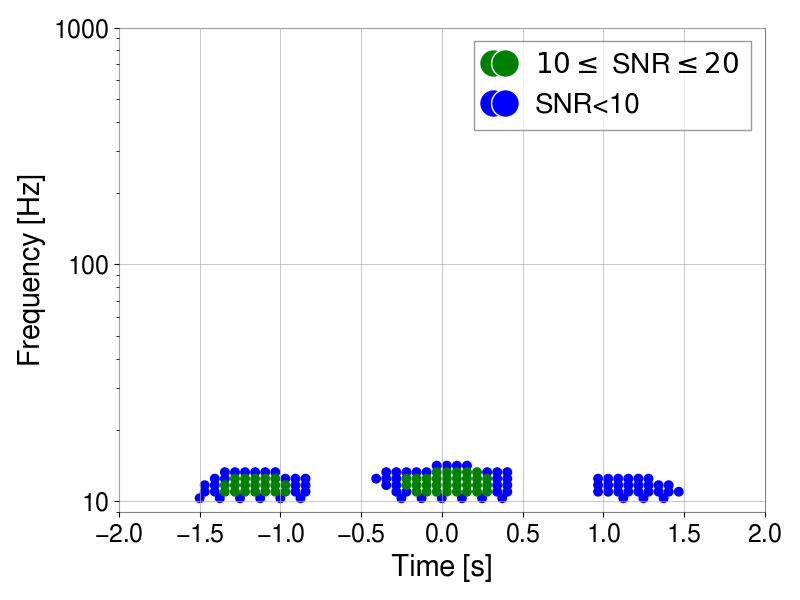}
        \label{fig:q16_23}
    \end{subfigure}
    \medskip
    \begin{subfigure}{0.32\textwidth}
        \centering
        \caption{$Q$-value = 30.64}
        \includegraphics[width=\textwidth]{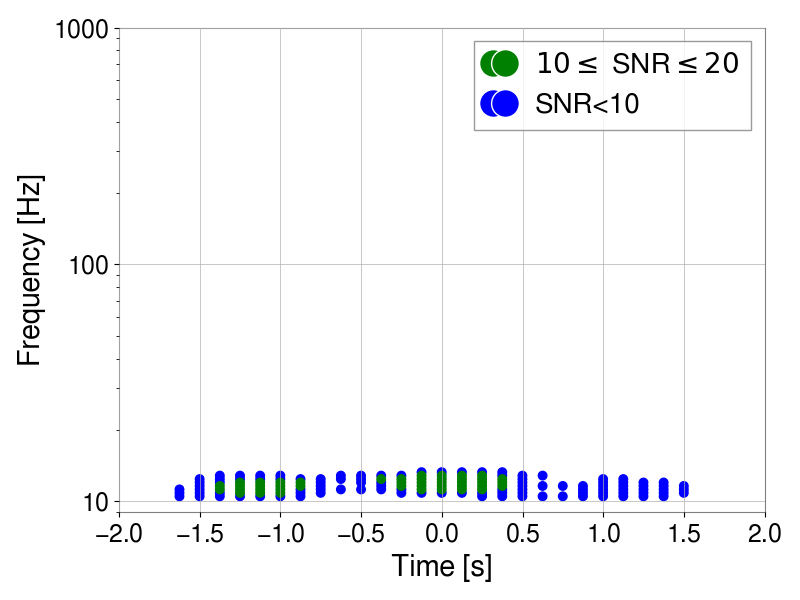}
        \label{fig:q30_64}
    \end{subfigure}
    \begin{subfigure}{0.32\textwidth}
        \centering
        \caption{$Q$-value = 57.84}
        \includegraphics[width=\textwidth]{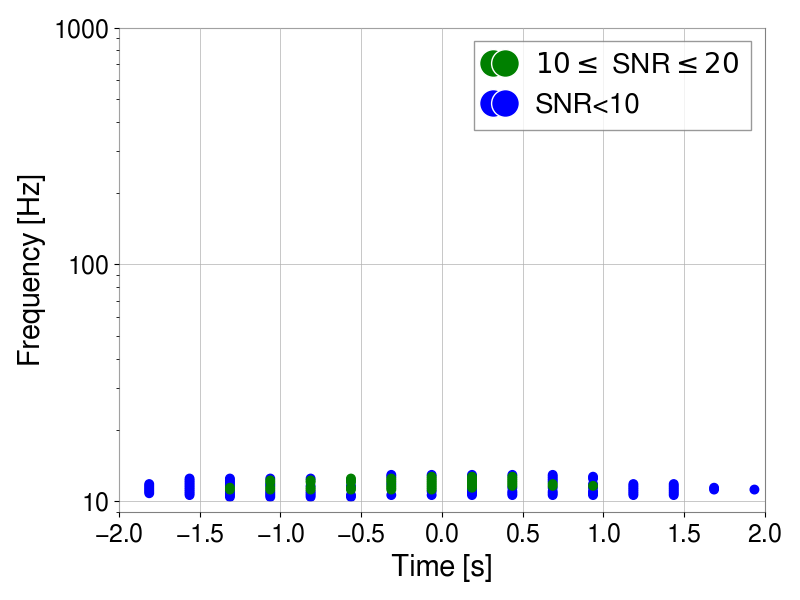}
        \label{fig:q57_84}
    \end{subfigure}
    \begin{subfigure}{0.32\textwidth}
        \centering
        \caption{Sum of all $Q$-values}
        \includegraphics[width=\textwidth]{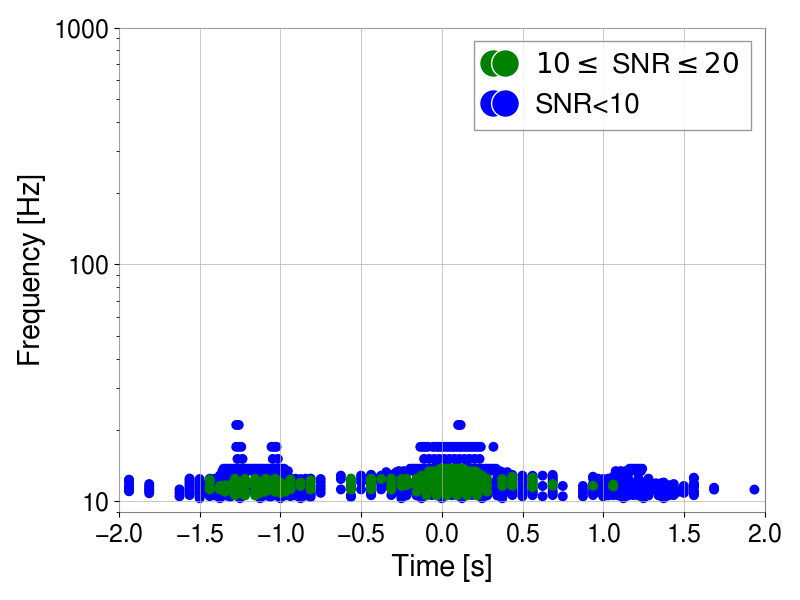}
        \label{fig:all_qs}
    \end{subfigure}
    \vspace{-0.5cm}
    \caption{Unclustered Omicron triggers during a low-frequency glitch at LLO. Panels (a) to (e) represent the same glitch at different $Q$-values, ranging from 4.56 to 57.84, illustrating how the glitch morphology changes with varying $Q$-values. While we maintain good time resolution with $Q$-values ranging from 4 to 16, we begin to lose that for values higher than 30 (d), transitioning towards more frequency resolution. Panel (f) shows all Omicron triggers for various $Q$-values; it can be regarded as a sum of all other $Q$-planes.}
    \label{fig:omicron_unclustered_glitch}
\end{figure}

For each one of these glitches, we can access the ROOT files to get the unclustered information. Figure \ref{fig:omicron_unclustered_glitch} illustrates the time-frequency distribution of unclustered triggers\footnote{Here, we refer to each data point from unclustered Omicron as a trigger.} across various $Q$-planes for one of the most common glitches observed during O4a at the LIGO Livingston Observatory (LLO). Blue indicates SNR below 10 and green indicates SNR values between 10 and 20. The parameter $Q$, from Equation~\ref{eq:q_value}, can be interpreted as follows: a higher $Q$-value results in higher resolution in frequency, while a lower $Q$-value results in worse resolution in frequency (or better resolution in time~\cite{gabor1946theory}). This characteristic is observable when comparing the images. Figure~\ref{fig:all_qs} presents the morphology when all $Q$-values are considered.

\begin{figure}[ht!]
    \centering
    \begin{subfigure}[b]{0.48\linewidth}
        \centering
        \caption{$Q$ = 8.60}
        \includegraphics[width=\linewidth]{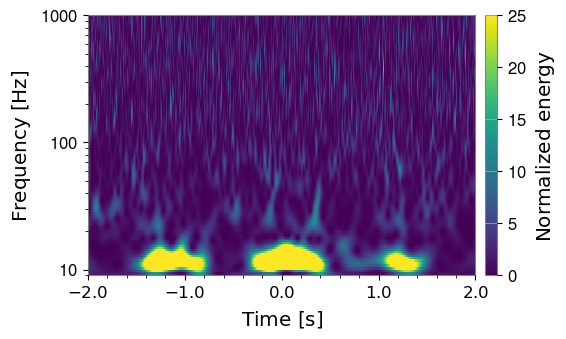}
        \label{fig:glitch_spectrogram_q8}
    \end{subfigure}\quad
    \begin{subfigure}[b]{0.48\linewidth}
        \centering
        \caption{$Q$ = 57.84}
        \includegraphics[width=\linewidth]{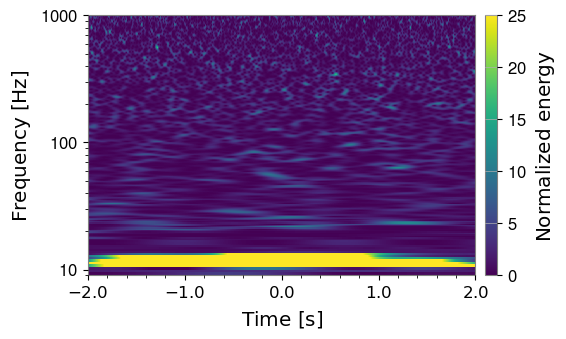}
        \label{fig:glitch_spectrogram_q57}
    \end{subfigure}
    \caption{Spectrograms with different $Q$-values generated with GWpy for the same glitch presented in~\Fref{fig:omicron_unclustered_glitch}. These subfigures illustrate the impact of varying $Q$-values on glitch morphology.}
    \label{fig:glitch_spectrogram}
\end{figure}

For comparison, Figure~\ref{fig:glitch_spectrogram} presents the spectrogram generated using the $Q$-transform via {\tt GWpy}~\cite{GWpy} for the same glitch at two different $Q$-values (8.60 and 57.84). This is one of the most common representations of transient signals in the detector's data. In Figure~\ref{fig:glitch_spectrogram_q8}, with a low $Q$-value, three repeating arches over time are visible, which is consistent with the first three images in Figure~\ref{fig:omicron_unclustered_glitch}. In contrast, in Figure~\ref{fig:glitch_spectrogram_q57}, the high $Q$-value reveals a prominent line, similar to the fourth and fifth images in Figure~\ref{fig:omicron_unclustered_glitch}.

\begin{figure}[ht!]
    \centering
    \begin{subfigure}{0.44\textwidth}
        \centering
        \caption{\textit{Scattered Light}}
        \includegraphics[width=\textwidth]{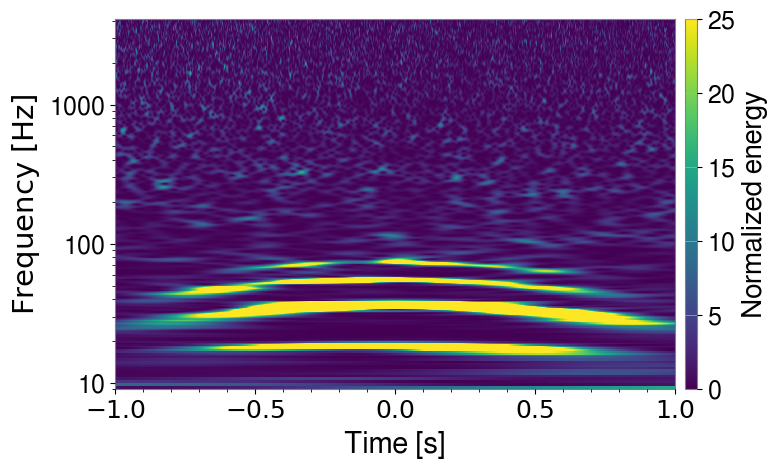}
        \label{fig:blip}
    \end{subfigure}
    \hspace{0.5cm}
    \begin{subfigure}{0.44\textwidth}
        \centering
        \caption{\textit{Tomte}}
        \includegraphics[width=\textwidth]{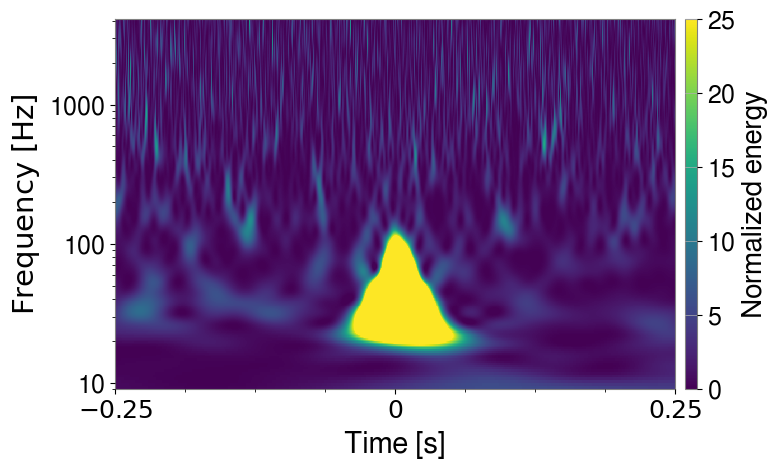}
        \label{fig:powerline}
    \end{subfigure}
     \medskip
    \begin{subfigure}{0.44\textwidth}
        \centering
        \caption{\textit{Blip}}
        \includegraphics[width=\textwidth]{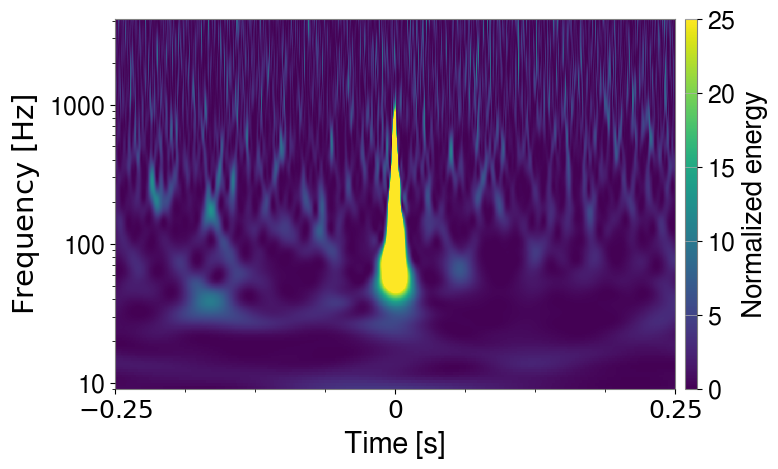}
    \end{subfigure}
    \hspace{0.5cm}
    \begin{subfigure}{0.44\textwidth}
        \centering
        \caption{\textit{Extremely Loud}}
        \includegraphics[width=\textwidth]{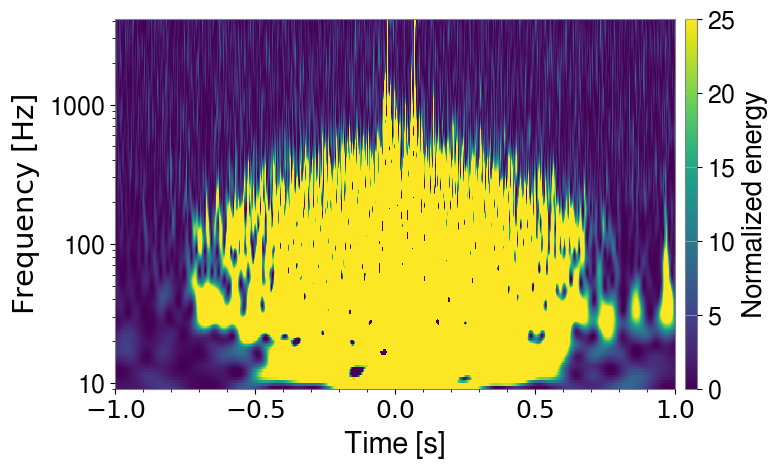}
    \end{subfigure}
    \caption{Spectrograms of four different classes of glitches selected during O3b at LLO. Panel (a): \textit{Scattered Light}, although many times these glitches are known to be caused by stray light, the main goal is to identify the surface that is scattering the light. Panel (b): \textit{Tomte}, observed during ESD bias voltage ramping \cite{42851}, with their exact causes still unknown. Panel (c): \textit{Blip}, a class of glitches with no identified sources~\cite{cabero2019blip}. Panel (d): \textit{Extremely Loud}, typically characterized by a very high SNR (greater than 100). Recent data suggest these glitches may be associated with scattered light near the End Test Mass (ETM)~\cite{alog73447,alog72968}.}
    \label{fig:classes_of_glitches}
\end{figure}

Glitches from the same source usually have similar features, allowing patterns to be identified in the data and enabling the creation of glitch classes. These features include arch shapes, their repetitions, the morphology and the width of bands (whether narrow or broad), SNR, frequency range (whether low or high), and others. Gravity Spy is a successful tool that involves volunteers, machine-learning techniques, and DetChar experts for classifying glitches~\cite{zevin2017gravity}. The names of the classes are derived either from their morphologies observed in the images or from prior knowledge of their sources. Examples include \textit{Tomte}, \textit{Low Frequency Burst}, \textit{Blip}~\cite{cabero2019blip}, \textit{Scattered Light}~\cite{soni2020reducing, soni2024modeling}, and \textit{Extremely Loud}~\cite{nichols2024investigations,alog73447,alog72968}, among others. Figure~\ref{fig:classes_of_glitches} presents spectrograms illustrating selected glitch classes observed during O3b at LLO. Further details on glitch classifications can be found in~\cite{zevin2017gravity,glanzer2023data,Zevin:2023rmt, GravitySpyZenodo}.

To classify glitches, tools like Gravity Spy rely on Omicron's clustered information. Whenever Omicron detects a glitch, these tools generate one or more images similar to Figure~\ref{fig:classes_of_glitches} and classify them accordingly. This paper analyzes glitches based on their morphologies derived from unclustered Omicron data using all $Q$-values, as shown in Figure~\ref{fig:all_qs}, enabling computationally efficient and rapid analysis without requiring the reapplication of the $Q$-transform or the creation of images as a training dataset.

The t-SNE (t-distributed stochastic neighbor embedding) algorithm is a statistical technique that allows visualization of high-dimensional data points in a lower-dimensional space~\cite{van2008visualizing}, typically two or three, which aligns with human visual perception capabilities. In addition, the algorithm promotes clustering based on the similarities between pairwise data points, offering an additional layer of interpretation for complex datasets. In our application, this visualization tool was implemented using scikit-learn~\cite{pedregosa2011scikit}, and some key hyperparameters - defined by the user - were evaluated, including \textit{perplexity} ($p$), \textit{learning\_rate}, \textit{init}, \textit{random\_state}, and \textit{n\_components}. The latter defines the dimensionality of the output space, which was set to 2 in our case.

The algorithm computes similarity probabilities in both the high- and low-dimensional spaces and adjusts the result by minimizing a cost function, the sum of Kullback–Leibler (KL) divergences, over all data points, such that the low-dimensional distribution resembles the high-dimensional one. The \textit{perplexity} is inside of this cost function and can be interpreted as a measure of the effective number of nearest neighbors, and it significantly influences the t-SNE output. According to~\cite{van2008visualizing}, typical perplexity values range from 5 to 50, although the optimal value may depend on dataset size and structure.

Since t-SNE is a visualization technique, some degree of human interpretation is required. In our case, we tested a broader range of perplexity values. The smallest value tested ($p=5$) produced less compact groupings. For the other values within the recommended range, the outputs were qualitatively similar. When testing larger values (above $p=70$) - although the main clusters remained noticeable - we observed visible distortions in the 2-dimensional output. Based on these observations, we selected a perplexity of $p=30$ for all visualizations presented in this work, as it corresponds to the function’s default value, falls within the recommended range, and provides a good balance between local and global structure.

The \textit{init} parameter defines how the algorithm initializes the low-dimensional embedding and can be set to either \textit{random} or \textit{pca} (from Principal Component Analysis). We tested both options and observed that distortions began to appear at similar values of $p$. To avoid introducing any initialization bias, we used \textit{random} for all visualizations presented in this work. Another important hyperparameter is the \textit{learning\_rate}, which affects the speed at which the layout stabilizes during optimization. It controls the step size of each iteration when minimizing the cost function. This parameter, along with others not explicitly discussed here, was used with its default value.

Finally, the \textit{random\_state} hyperparameter controls the seed for random number generation. If not set, t-SNE may produce different cluster arrangements across runs due to its stochastic nature. Fixing this parameter ensures reproducibility. To assess the stability of the embedding under the chosen configuration, we performed 100 runs using the same data and hyperparameters but with different random seeds, and recorded the KL divergence for each case. The resulting values were highly consistent, with a standard deviation of $0.0064$ and a coefficient of variation of only $0.5\%$, indicating low variability across runs. For the visualizations presented in this work, we used the embedding generated with \textit{random\_state} = 0. It is important to highlight that our main interest lies in visualizing the internal structure of the glitch population. The groups of glitches are the important information here, not their positions, which makes t-SNE an appropriate choice, as it preserves local similarities.

After configuring these parameters, the primary contribution of this study is to propose a complementary use of t-SNE for the characterization of glitches derived from Omicron features. This approach removes the need to create an image dataset for training machine learning models, significantly accelerating algorithmic analysis. Another important aspect is that this analysis can be conducted independently, without prior knowledge of glitch classes. In the following section, we begin by using the Gravity Spy classes as a reference to investigate how glitch duration, SNR normalization, and the exclusive use of either low or high $Q$-values influence the results.

\section{t-SNE Applied to LLO Glitches During O3b with Prior Knowledge of Classes}
\label{sec:method_with_classes}
We selected a dataset of $1,000$ random glitches from each of the ten most common Gravity Spy classes in O3b (with a minimum of $90\%$ confidence in each glitch's classification). A representation, as shown in the last image of \Fref{fig:omicron_unclustered_glitch}, was built using all $Q$-values. Each image was then divided into $30\times41$ grid of pixels that could include zero, one, or more omicron data points (triggers). Each pixel stored the value of the trigger with the highest SNR. Using this data, a matrix was created for each transient, which was then flattened into a data vector of $1,230$ elements. This approach is similar to the one discussed in~\cite{ferreira2022comparison}. However, it differs by using shorter-duration intervals in the creation of glitch morphology, higher time resolution, and normalized SNR values. More details will be discussed in the following sections. 

\begin{figure}[ht!]
    \centering
    \includegraphics[width=1\textwidth]{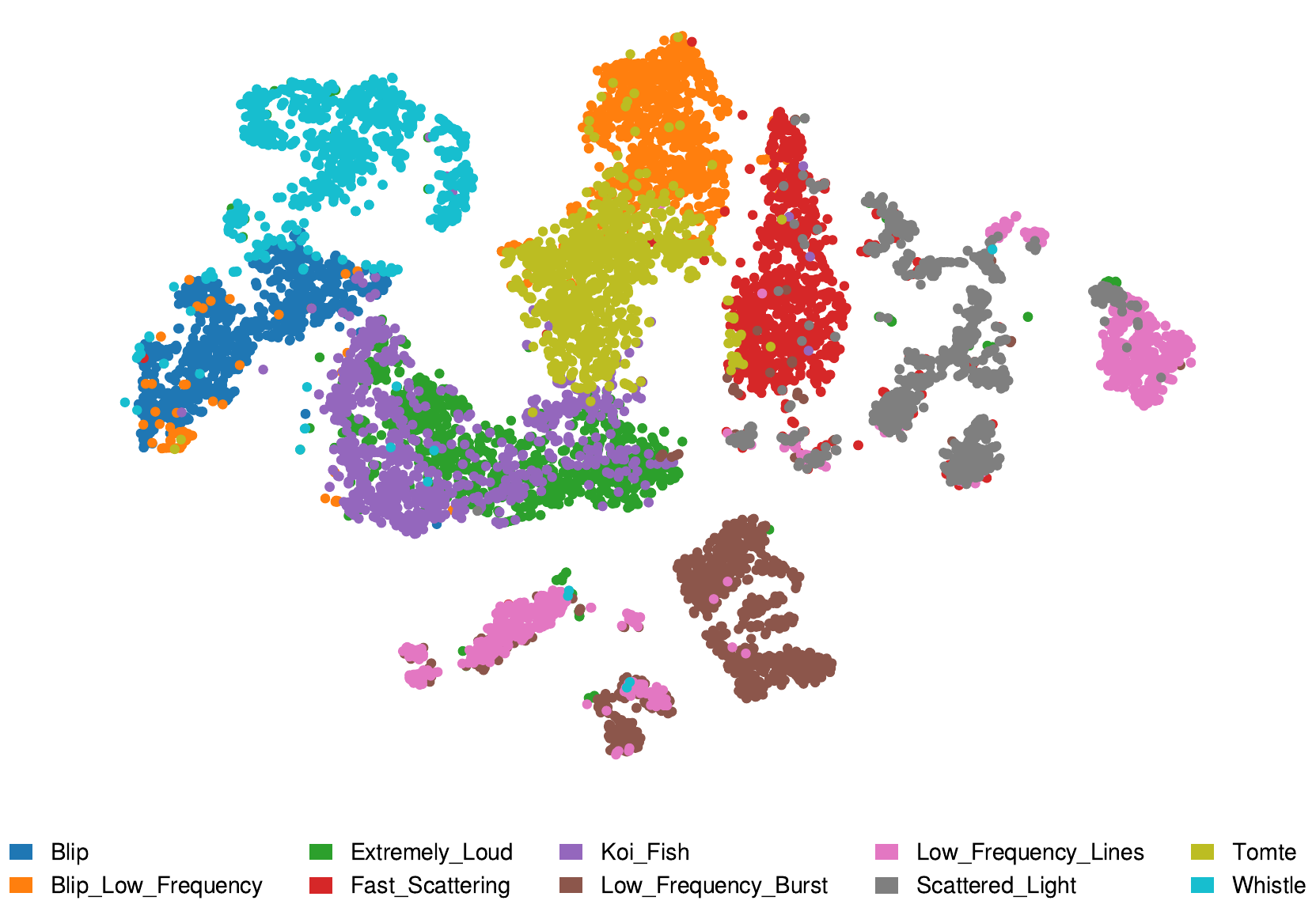}
    \caption{Scatter plot of t-SNE coordinates representing 10,000 data points from a 1,230-dimensional space derived from unclustered Omicron triggers. Each point represents a glitch, with colors indicating the classification according to Gravity Spy. The input data included triggers from all $Q$-planes, as depicted in Figure~\ref{fig:all_qs}.}
    \label{fig:o3b_tsne_l1}
\end{figure}

Figure~\ref{fig:o3b_tsne_l1} presents a scatter plot of the two t-SNE output coordinates applied to this dataset, where each color represents a different Gravity Spy class. Overall, the groups are well clustered - that is, the clusters align with the labels - indicating that the vector representation effectively characterizes the transients. It is important to note that the t-SNE coordinates are used for visualizing the data; they do not have physical meaning.

The class \textit{Low Frequency Lines} (in pink) shows two distinct subgroups. By analyzing the frequency content of the transients in each subgroup, we find that one, located on the right side, contains glitches with frequencies around $\SI{20}{Hz}$ and the other, located in the center bottom region, around $\SI{12}{Hz}$ and $\SI{13}{Hz}$. Additionally, a small group exists between this last subgroup and the \textit{Low Frequency Burst} class (in brown), with a frequency around $\SI{11}{Hz}$. These distinctions are related to the frequency resolution used in constructing the dataset and were identified by inspecting the glitch properties, not inferred from the spatial arrangement in the t-SNE plot. A similar subdivision appears within the \textit{Fast Scattering} class (in red), where one group of glitches is centered around $\SI{27}{Hz}$ (corresponding to the lower subdivision of the group), and another around $\SI{39}{Hz}$ (seen in the upper subdivision).

\begin{figure}[ht!]
    \centering

    \begin{subfigure}{0.48\textwidth}
        \centering
        \includegraphics[width=\textwidth]{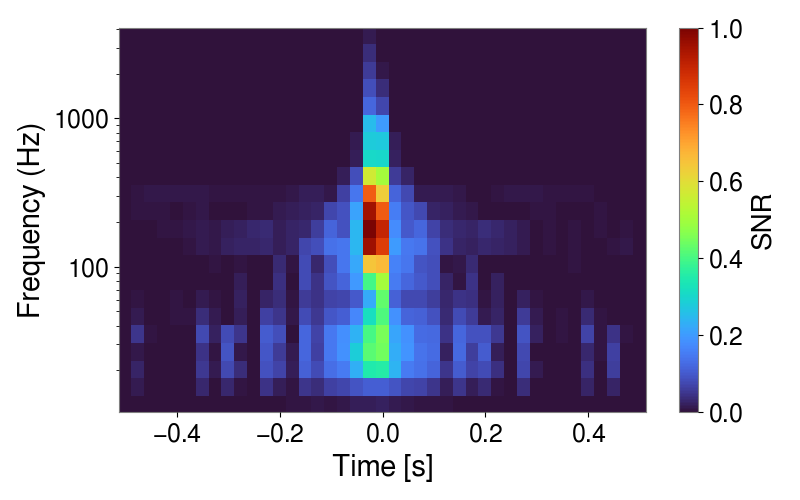}
        \caption{\textit{Extremely Loud} \\ Created from unclustered Omicron data}
        \label{fig:ext_matrix}
    \end{subfigure}
    \hfill
    \begin{subfigure}{0.48\textwidth}
        \centering
        \includegraphics[width=\textwidth]{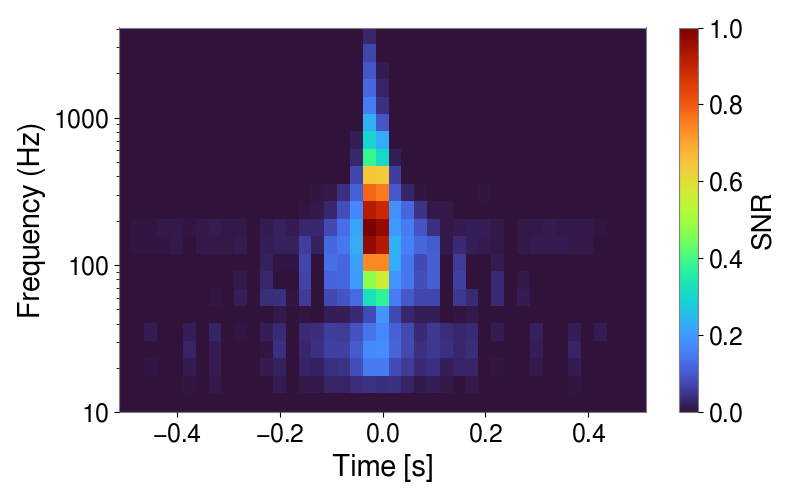}
        \caption{\textit{Koi Fish} \\ Created from unclustered Omicron data}
        \label{fig:koifish_matrix}
    \end{subfigure}

    \vspace{0.3cm}

    \begin{subfigure}{0.48\textwidth}
        \centering
        \includegraphics[width=\textwidth]{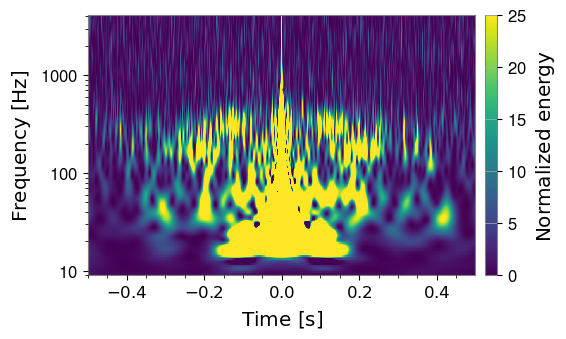}
        \caption{\textit{Extremely Loud} \\ \texttt{GWpy} spectrogram with $Q=5.66$}
        \label{fig:spec_extloud_lowq}
    \end{subfigure}
    \hfill
    \begin{subfigure}{0.48\textwidth}
        \centering
        \includegraphics[width=\textwidth]{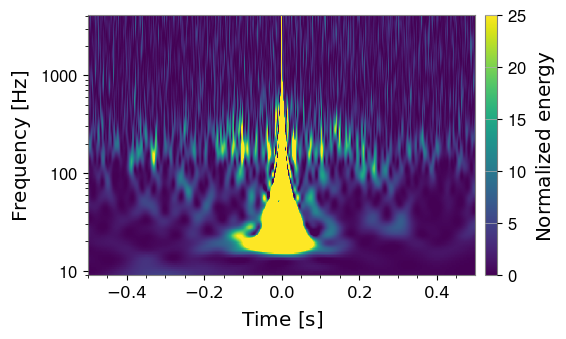}
        \caption{\textit{Koi Fish} \\ \texttt{GWpy} spectrogram with $Q=5.66$}
        \label{fig:spec_koifish_lowq}
    \end{subfigure}

    \vspace{0.3cm}

    \begin{subfigure}{0.48\textwidth}
        \centering
        \includegraphics[width=\textwidth]{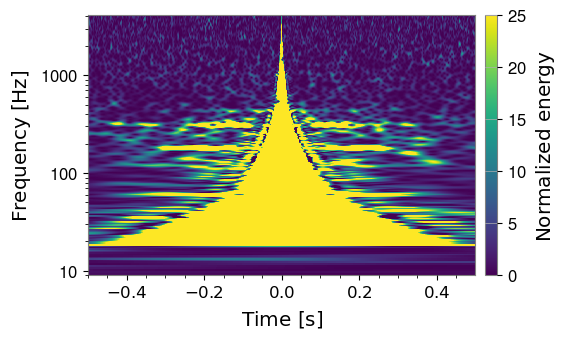}
        \caption{\textit{Extremely Loud} \\ \texttt{GWpy} spectrogram with $Q=36.26$}
        \label{fig:spec_extloud_highq}
    \end{subfigure}
    \hfill
    \begin{subfigure}{0.48\textwidth}
        \centering
        \includegraphics[width=\textwidth]{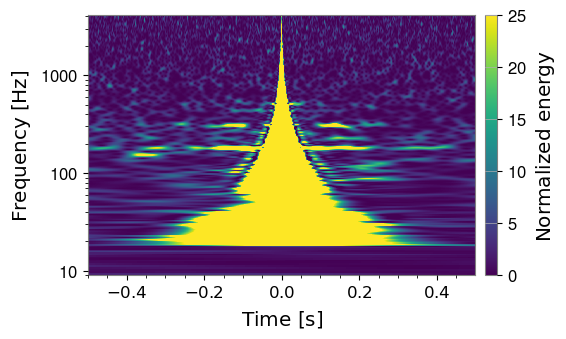}
        \caption{\textit{Koi Fish} \\ \texttt{GWpy} spectrogram with $Q=36.26$}
        \label{fig:spec_koifish_highq}
    \end{subfigure}

    \caption{Two glitches classified as \textit{Extremely Loud} (a) and \textit{Koi Fish} (b), represented by data matrices from unclustered Omicron data used as input to t-SNE. SNR normalization highlights the morphological similarity between them, which helps explain the overlap observed in the t-SNE output. Their representations in spectrograms generated with \texttt{GWpy} using low (c,d) and high (e,f) $Q$-values are included for comparison.}
    \label{fig:ext_vs_koifish}
\end{figure}

There is a significant overlap between \textit{Koi Fish} (purple) and \textit{Extremely Loud} (green) classes.
Figures~\ref{fig:ext_matrix} and~\ref{fig:koifish_matrix} show two data matrices, generated from unclustered Omicron files and used as input to t-SNE, containing glitches classified as \textit{Extremely Loud} and \textit{Koi Fish}, respectively. Indeed, some of them are morphologically very similar, and the main feature that differentiates them is the SNR, as will be shown in subsection~\ref{subsubsec:influence of SNR}; further details about their similarities are also discussed in~\cite{nichols2024investigations}.

For comparison, Figures~\ref{fig:spec_extloud_lowq} and~\ref{fig:spec_extloud_highq} show the same \textit{Extremely Loud} glitch, plotted using \texttt{GWpy} with low and high $Q$-values, respectively. Corresponding plots for the selected \textit{Koi Fish} glitch are shown in Figures~\ref{fig:spec_koifish_lowq} and~\ref{fig:spec_koifish_highq}. In both cases, the selected $Q$-plane from \texttt{GWpy}, which contains the tile with the highest SNR, corresponds to the lowest $Q$-value.

As can be seen, the images generated from unclustered Omicron data (Figures~\ref{fig:ext_matrix} and~\ref{fig:koifish_matrix}) display a stronger vertical central region, which is also prominent in the spectrograms with low $Q$-values (Figures~\ref{fig:spec_extloud_lowq} and~\ref{fig:spec_koifish_lowq}). However, since the matrices represent the sum across all $Q$-planes, as previously discussed, a weaker contour around this central region is also visible. This secondary structure can be observed more clearly in the spectrograms with higher $Q$-values. As mentioned earlier, all similar glitches are affected in a comparable way.

Another important aspect to mention is that the \textit{Scattered Light} class was subdivided into smaller groups. This subdivision occurred because the technique is sensitive to variations in the duration of the arches, the number of arches, and the peak frequency.

\subsection{Influence of window duration:}

\begin{figure}
    \centering
    \begin{subfigure}{0.49\textwidth}
        \centering
        \caption{All $Q$-values - Duration: 4s}
        \includegraphics[width=\textwidth]{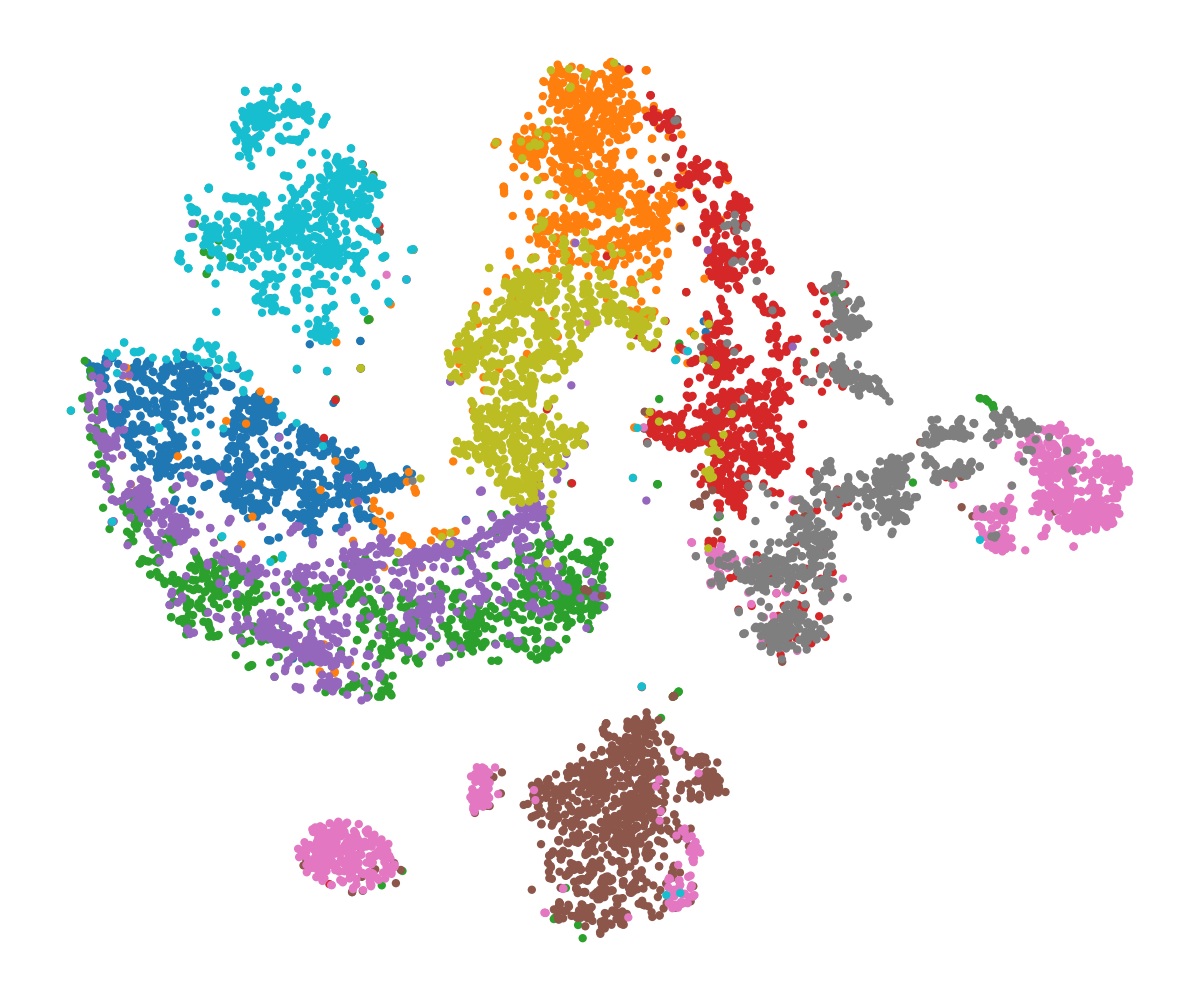}
        \label{fig:tsne_2sec}
    \end{subfigure}
    \begin{subfigure}{0.49\textwidth}
        \centering
        \caption{All $Q$-values - SNR not normalized}
        \includegraphics[width=1\textwidth]{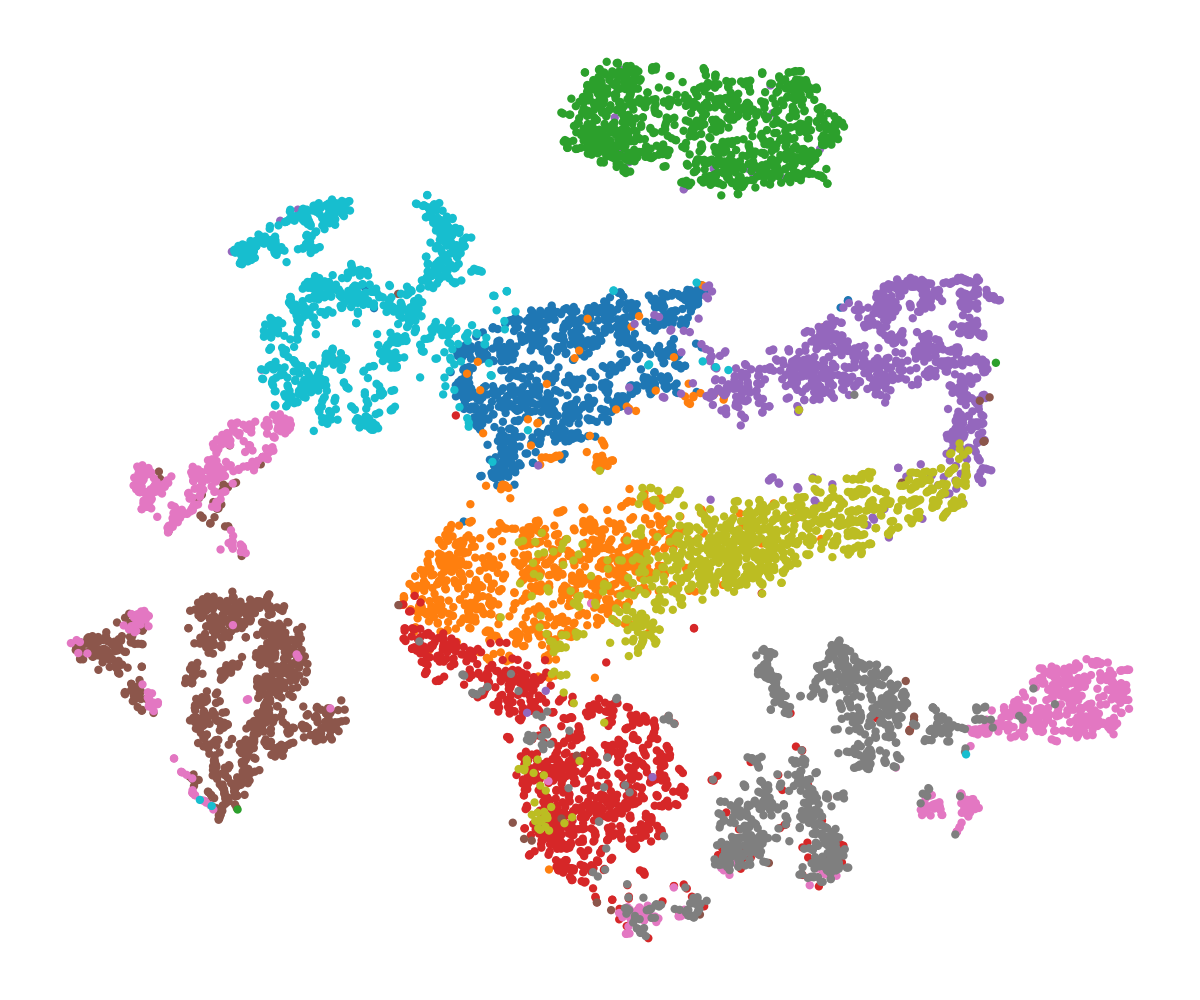}
        \label{fig:tsne_including_loudglitches}
    \end{subfigure}
    \medskip
        \begin{subfigure}{0.49\textwidth}
        \centering
        \caption{$Q$-values $\leq$ 10}
        \includegraphics[width=\textwidth]{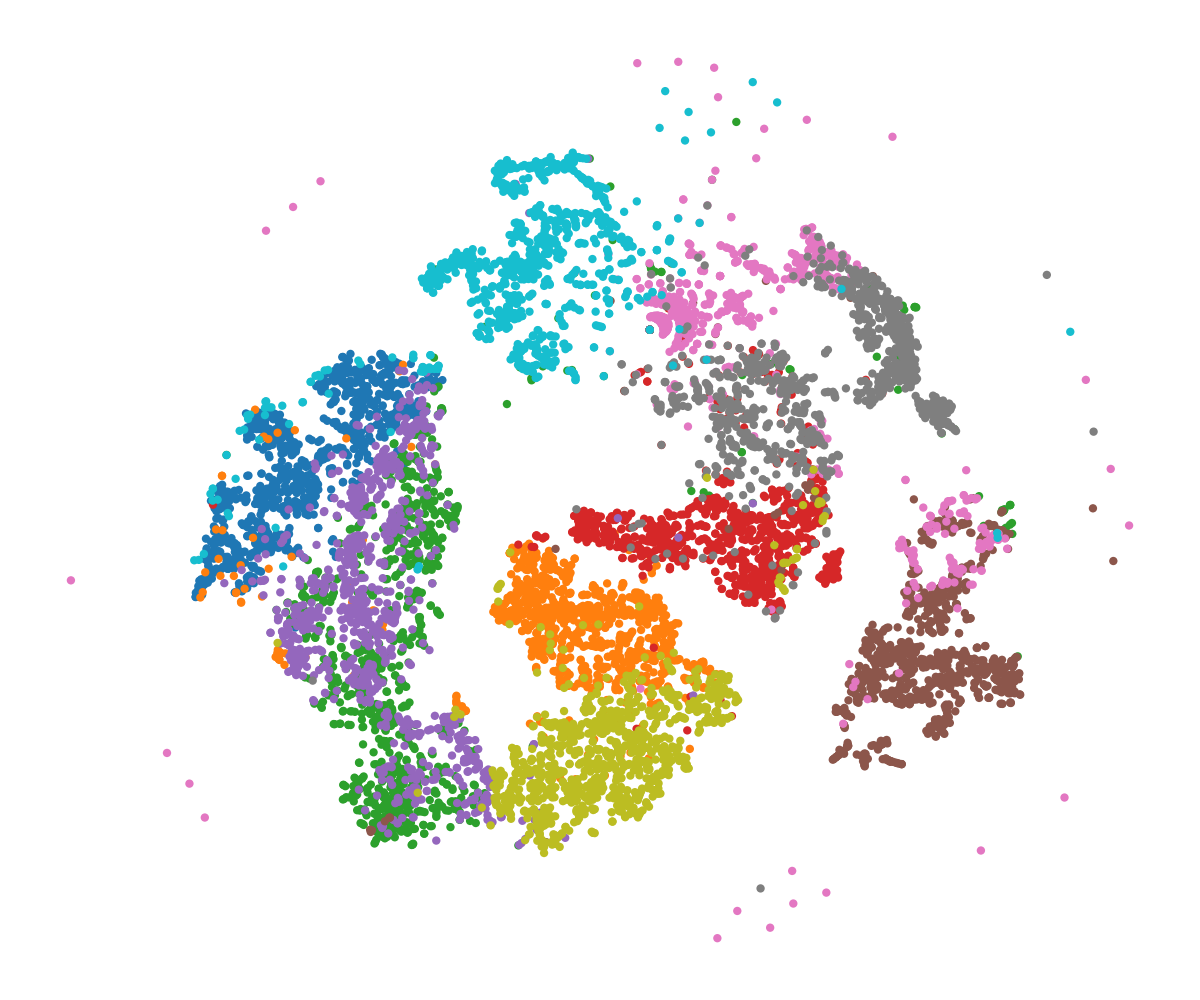}
        \label{fig:tsne_low_Q}
    \end{subfigure}
    \begin{subfigure}{0.49\textwidth}
        \centering
        \caption{$Q$-values $\geq$ 10}
        \includegraphics[width=1\textwidth]{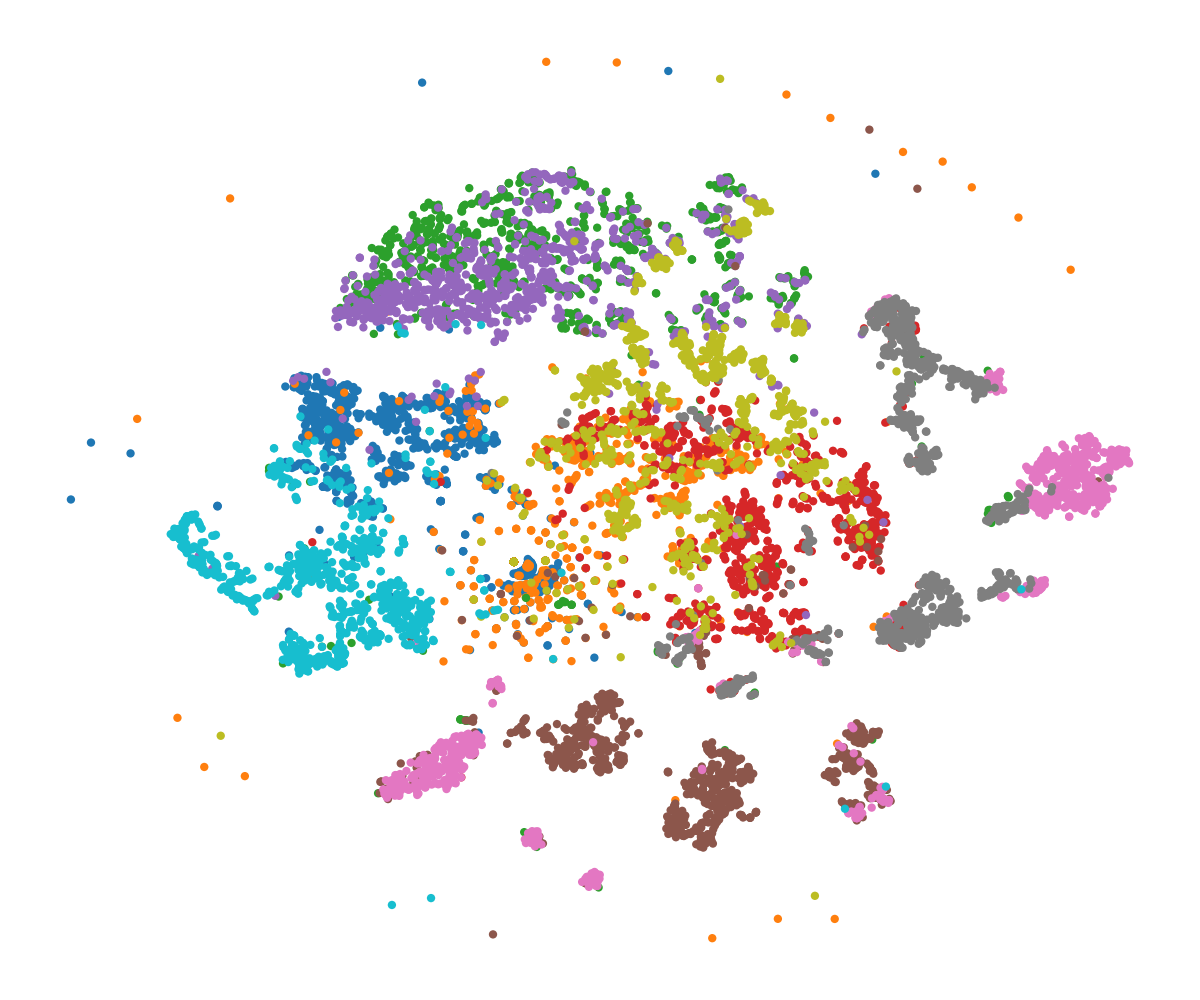}
        \label{fig:tsne_high_Q}
    \end{subfigure}
    \medskip
    \begin{subfigure}{1\textwidth}
        \centering
        \includegraphics[width=1\textwidth]{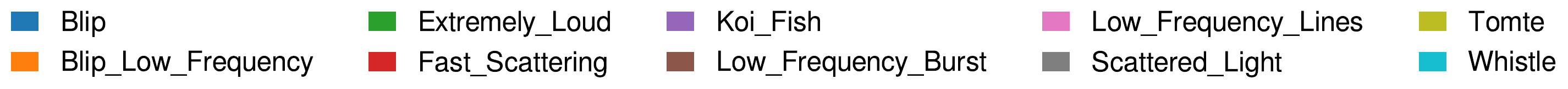}
        \label{fig:legend}
    \end{subfigure}
    \caption{Effects on the t-SNE output for the same dataset used in Figure~\ref{fig:o3b_tsne_l1}, with small changes in input parameters. (a) 4-second window instead of 1-second: although groups of colors are visible, they appear more interconnected. (b) Same dataset (with 1-second window) using non-normalized SNR: the \textit{Extremely Loud} class (green) stands out due to its high SNR, forming an isolated cluster, while other classes appear more interconnected. (c) Dataset built from unclustered Omicron files filtered to include only $Q$-values less than 10: this illustrates the impact of $Q$-value selection. (d) Triggers with $Q$-values greater than 30: higher $Q$-values result in more overlap in the center. Each $Q$-range can be selected depending on the desired resolution in time or frequency.}
    \label{fig:tsne_changing_parameters}
\end{figure}

When a glitch is centered at time $t$ and a window duration $w$ is selected, triggers are sought within the time interval $t - w/2$ to $t + w/2$. The current primary classification method used in LIGO data, Gravity Spy, employs four different image windows to classify each glitch: $\SI{0.5}{s}$, $\SI{1.0}{s}$, $\SI{2.0}{s}$, and $\SI{4.0}{s}$~\cite{zevin2017gravity, wu2024advancing}, allowing for large-scale analysis as well as examination of finer details. Figure~\ref{fig:tsne_2sec} presents the t-SNE output when we use $w = \SI{4.0}{s}$. Although it's still possible to see clusters of colors, there are more overlaps than before in Figure~\ref{fig:o3b_tsne_l1}, where a window duration of $w = \SI{1.0}{s}$ is used. This increased overlap occurs because longer durations make it more difficult to distinguish short-duration glitches, effectively causing a ``zoom out" effect on the morphological features, which explains why \textit{Blip}, \textit{Koi Fish}, and \textit{Extremely Loud} glitches appear closer together. Although shorter durations provide clearer distinctions between some classes, we opt for keeping $w = \SI{1.0}{s}$, which allows us to observe arches (as shown in Figure~\ref{fig:classes_of_glitches}), but not necessarily the repetition of them.

\subsection{Influence of \rm{SNR}}
\label{subsubsec:influence of SNR}
The proposed method exhibits sensitivity to the normalization of the data vectors. Specifically, for each vector, the elements ($\text{SNRs}$) are normalized using the minimum (\( \text{SNR}_{\min} \)) and maximum (\( \text{SNR}_{\max} \)) SNR values of that vector, resulting in the normalized elements \( \text{SNRs}_{\text{norm}} \), given by 

\begin{center}
\begin{equation}
\text{SNRs}_{\text{norm}} = \frac{\text{SNRs} - \text{SNR}_{\min}}{\text{SNR}_{\max} - \text{SNR}_{\min}},
\end{equation}
\end{center}

with \( \text{SNRs}_{\text{norm}} \) varying from 0 to 1.

Figure~\ref{fig:tsne_including_loudglitches} presents the same analysis performed without normalizing SNR values. This lack of normalization can assist in identifying outliers and stronger glitches in the dataset. However, as morphology is prioritized as the primary distinguishing feature among classes in this study, we utilize normalized data as a reference. In the unnormalized application, the \textit{Extremely Loud} class is now distinctly isolated and no longer mixed with \textit{Koi Fish} class, unlike the observation in Figure~\ref{fig:o3b_tsne_l1}. Although some well-defined clusters are still observable, a drawback of this approach is that the well-positioned cluster tends to ``repel'' other groups, making them appear more interconnected. This results in a division between regions with high SNR glitches (comprised solely of \textit{Extremely Loud} in this case) and those with lower SNR glitches (containing all remaining classes).

\subsection{Influence of $Q$-Values}

\begin{figure}[ht!]
    \centering
    \begin{subfigure}{0.44\textwidth}
        \centering
        \caption{Glitch 1 with a $Q$-value of 5.66}
        \includegraphics[width=\textwidth]{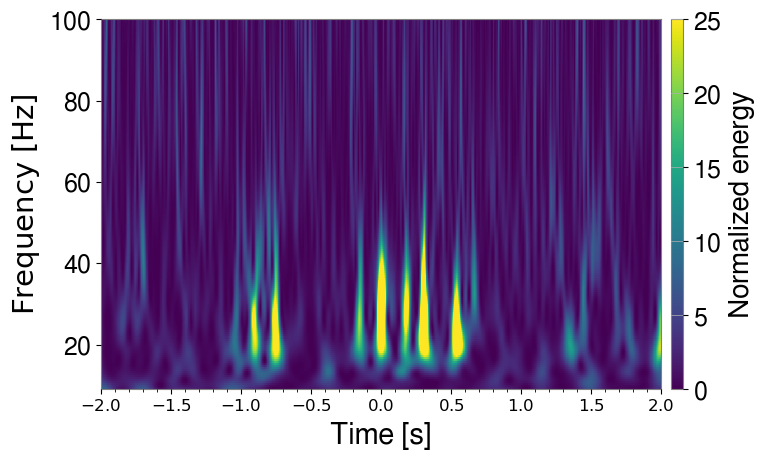}
        \label{fig:h1_im1_lowQ}
    \end{subfigure}
    \hspace{0.5cm}
    \begin{subfigure}{0.44\textwidth}
        \centering
        \caption{Glitch 1 with a $Q$-value of 45.25}
        \includegraphics[width=\textwidth]{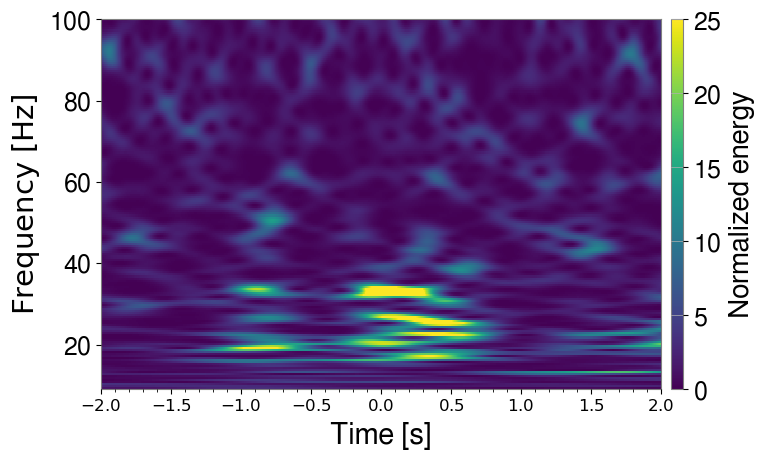}
        \label{fig:h1_im1_highQ}
    \end{subfigure}
     \medskip
    \begin{subfigure}{0.44\textwidth}
        \centering
        \caption{Glitch 2 with a $Q$-value of 45.25}
        \includegraphics[width=\textwidth]{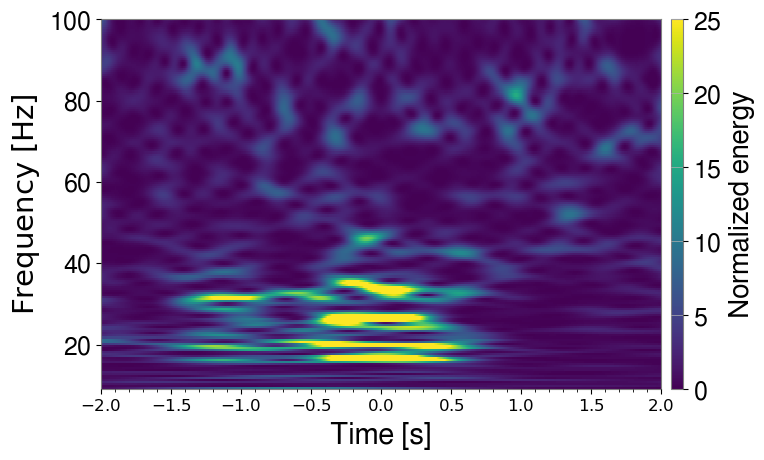}
        \label{fig:h1_im2_highQ}
    \end{subfigure}
    \hspace{0.5cm}
    \begin{subfigure}{0.44\textwidth}
        \centering
        \caption{Glitch 2 with a $Q$-value of 5.66}
        \includegraphics[width=\textwidth]{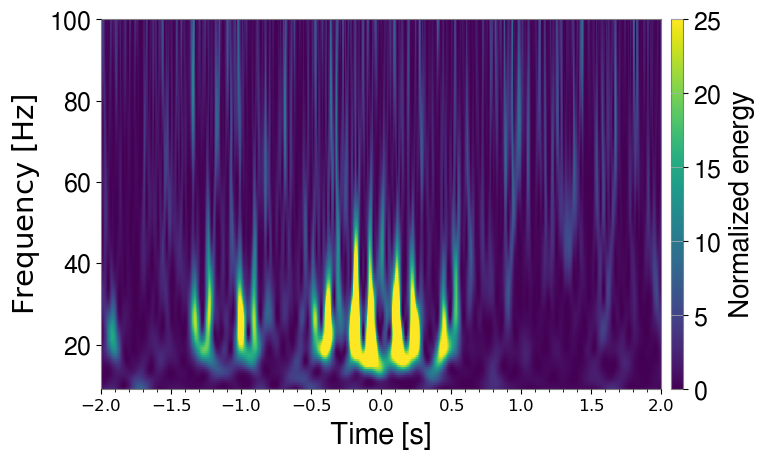}
        \label{fig:h1_im2_lowQ}
    \end{subfigure}
    \caption{Visualization of glitches in the broadband frequency range from 20 to \SI{40}{\Hz} at LHO during the O4a run. The spectrograms show two glitches, (a) and (c), that appear distinct but actually originate from the same source. Their respective spectrograms with different $Q$-values, (b) and (d), illustrate how the choice of $Q$-value can influence glitch morphology and potentially lead to misclassification and/or misinterpretation.}
    \label{fig:o4a_diff_qvalues}
\end{figure}

The $Q$-value is intrinsically linked to the time and frequency resolutions of the glitch spectrogram. This relationship implies that, in certain cases, the choice of $Q$-value can significantly influence the morphology of the glitch, as demonstrated in Figures~\ref{fig:omicron_unclustered_glitch} and~\ref{fig:glitch_spectrogram}. During the O4a run, LHO was notably impacted by glitches in the broadband frequency range from $20$ to \SI{40}{\Hz}~\cite{soni2024ligo} (as also presented in Figure~\ref{fig:omicron_clustered_day}). An examination of the spectrograms revealed that some of them exhibited repeated vertical patterns (Figure~\ref{fig:h1_im1_lowQ}), while other presented horizontal structures (Figure~\ref{fig:h1_im2_highQ}).

In the first glitch (Figure~\ref{fig:h1_im1_lowQ}), the $Q$-value is $5.66$, selected directly from \texttt{GWpy}, which determines the $Q$-value based on the $Q$-plane containing the tile with the highest SNR. The highest SNR plane for the second glitch in Figure~\ref{fig:h1_im2_highQ} corresponds to a $Q$-value of $45.25$. Despite their distinct appearances, these glitches have been identified as originating from the same source~\cite{71005,71092,71927}. To further investigate this effect, we analyzed each glitch using the $Q$-value originally assigned to the other. Figures~\ref{fig:h1_im1_highQ} and \ref{fig:h1_im2_lowQ} show the first glitch with the higher $Q$-value ($45.25$) and the second with the lower $Q$-value ($5.66$), respectively. As observed, although these glitches may initially appear different, their similarities become evident when analyzed using the same $Q$-value.

Figures~\ref{fig:tsne_low_Q} and~\ref{fig:tsne_high_Q} illustrate the impact of $Q$-values on t-SNE clustering when morphologies are derived from low $Q$-values (below 10) and high $Q$-values (greater than 30), respectively. Both scenarios are affected by outliers on the edges, forming a circular arrangement. The output using high $Q$-values exhibits more overlaps in the central region among different classes and also results in some classes being subdivided into additional subclasses, as for the \textit{Low Frequency Burst} (in brown), which reveals classes with different frequencies. Here, we highlight a potential application of this technique: selecting the $Q$-value range according to specific interests. For instance, we can study a single class with high $Q$-values to explore possible subclasses - like the distinct frequencies seen in \textit{Low Frequency Burst} - or focus on low $Q$-values to analyze longer durations and detect repetition over time. 

In our approach, we do not select a single optimal $Q$-value; instead, we utilize all available $Q$-values. This comprehensive approach enables us to capture all relevant information. While this method does incorporate the limitations associated with varying resolutions in time and frequency, glitches originating from the same source will be similarly affected, resulting in comparable morphologies. This consistency is beneficial compared to selecting only one $Q$-value, which may not always capture the full range of morphological features.

\subsection{Applying a classifier}
\label{sec:classifier}
Supervised machine learning techniques typically require a training dataset to predict the class of a new glitch. While these methods can perform very well, building an effective training dataset presents challenges, especially when new data classes, such as previously unknown glitches, may emerge. The previous study by~\cite{ferreira2022comparison} proposes a classification approach based on the Support Vector Machine technique~\cite{noble2006support}, which learns from previously labeled data. Here, we present an alternative method for glitch classification that is entirely independent of prior knowledge about the classes or their number.

The approach involves determining the number of clusters from the t-SNE output and assigning classes based on its two-dimensional representation.
Although t-SNE does not preserve the global structure of the data, this dimensionality reduction method captures and visualizes local relationships in high-dimensional space, enabling the identification of clusters in a low-dimensional representation. This is illustrated in Figure~\ref{fig:o3b_tsne_l1}, where t-SNE reveals clusters of glitches in the gravitational wave channel. The classifier presented here is entirely based on the two t-SNE coordinates, which do not have a direct physical meaning. The main goal is to identify groups of similar glitches and, based on the positions of these groups, define corresponding classes.

To achieve this, we applied {\tt Agglomerative Clustering}, a hierarchical clustering method that assigns classes to datasets~\cite{xu2005survey}. The algorithm begins by considering each data point as an individual cluster and iteratively merges pairs of clusters into a single cluster based on the configuration that minimizes the total within-cluster variance, calculated as the sum of squared Euclidean distances between each point and the centroid of its cluster. This approach is also referred to as Ward's method~\cite{ward1963hierarchical,sharma2019comparative}. This merging process continues until a stopping criterion is reached, which, in our case, was limited to a maximum of 30 clusters. Agglomerative Clustering generated, therefore, a clustering configuration for each number of clusters in the range from 2 to 30. 

The best configuration was determined by evaluating these different numbers of clusters using the {\tt Silhouette Score}~\cite{rousseeuw1987silhouettes}, which evaluates the quality of the cluster configuration\footnote{Both methods were implemented using the Python library {\tt Scikit-learn}~\cite{pedregosa2011scikit}.}. For each data point $i$, the Silhouette Score calculates the mean Euclidean distance $a$ between $i$ and all other data points in the same cluster. It also computes the distance $b$ between $i$ and data points from the nearest cluster, where the nearest cluster is defined as the one with the smallest mean Euclidean distance between $i$ and all points in the clusters to which $i$ does not belong. The score $s(i)$ is, therefore, given by

\begin{equation}
    s(i) = \frac{b(i) - a(i)}{\max(a(i), b(i))}.
\end{equation} 

\begin{figure}[ht!]
    \centering
    \begin{subfigure}[b]{0.485\textwidth}
        \centering
        \caption{Normalized data}
        \includegraphics[width=\textwidth]{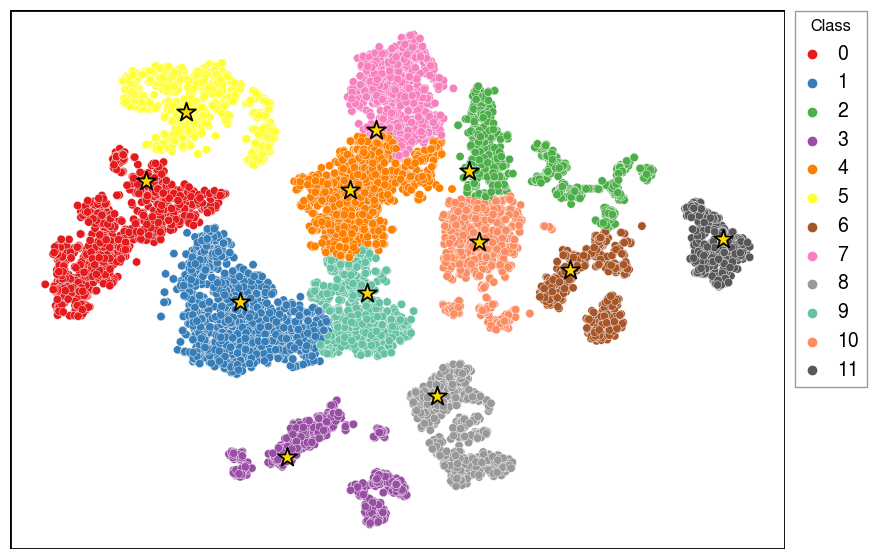}
        \label{fig:classifier_output_norm}
    \end{subfigure}\quad
    \begin{subfigure}[b]{0.485\textwidth}
        \centering
        \caption{Non-normalized data}
        \includegraphics[width=\textwidth]{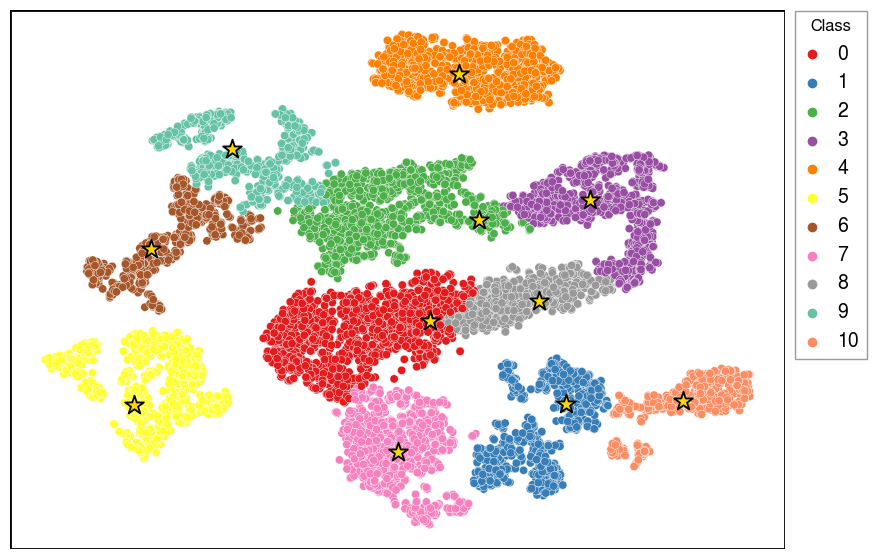}
        \label{fig:classifier_output_notnorm}
    \end{subfigure}
    \caption{The same t-SNE output, now colored according to the class assignments obtained via Agglomerative Clustering, for (a) normalized data and (b) non-normalized data. Stars indicate the glitches that are closest (in the original high-dimensional space) to the representative vector of each class. For comparisons with Gravity Spy classes, refer to Figure~\ref{fig:o3b_tsne_l1} and Figure~\ref{fig:tsne_including_loudglitches}, respectively.}
    \label{fig:classifier_output}
\end{figure}

The score ranges from -1 to 1, where a value of $1$ indicates the best configuration. The optimal number of clusters is then selected based on the highest mean Silhouette Score, indicating the best fit for the output data. Using the same dataset from Figure~\ref{fig:o3b_tsne_l1}, the Silhouette Score evaluated the clustering configurations generated by Agglomerative Clustering and indicated 12 as the optimal number of classes for this dataset. Figure~\ref{fig:classifier_output_norm} presents the t-SNE output, colored according to this classification method. The classes are now labeled from `'Class 0" to ``Class 11".

In comparison to Figure~\ref{fig:o3b_tsne_l1} and the Gravity Spy classes, our method identified two additional classes. One of these, depicted in purple as `Class 3', corresponds to glitches classified as \textit{Low Frequency Lines} and \textit{Low Frequency Burst}. As discussed in Section~\ref{sec:method_with_classes}, new groups were anticipated due to the sensitivity of t-SNE to different frequency ranges. This emphasizes that frequency resolution can be adjusted according to specific interests as well. Here, we are interested in presenting groups with differing peak frequencies, as they may originate from distinct sources.

The second new class, shown in green and designated as `Class 2', differentiates between glitches within the scattering categories. This class includes glitches classified as \textit{Fast Scattering} by Gravity Spy (indicated in red in Figure~\ref{fig:o3b_tsne_l1}) and \textit{Scattered Light} (in gray). Analysis of spectrograms for both types revealed that, although they do not exhibit similar morphology, they tend to have higher frequencies compared to those in the main clusters of \textit{Scattered Light} and \textit{Fast Scattering}.

Figure~\ref{fig:classifier_output_notnorm} presents the results obtained without normalizing SNR, as illustrated in Figure~\ref{fig:tsne_including_loudglitches}. In this case, since most glitches are more interconnected due to loud glitches, close proximities may increase classification errors. However, the class labels generally align with Gravity Spy in both cases, indicating that data normalization can be applied based on specific analysis goals.

After determining the optimal number of classes and their distribution, we developed a classification method for new glitches. A representative vector was generated for each class using the 1,230 dimensions from unclustered Omicron data: $80\%$ of each group’s data was used to compute the mean vector, while the remaining $20\%$ served as a test dataset. In Figure~\ref{fig:o3b_tsne_l1}, a star marks the glitch that is closest - within the high-dimensional space - to the representative vector of each class (computed from 80\% of the data).

\begin{figure}[ht!]
    \centering
    \includegraphics[width=0.92\textwidth]{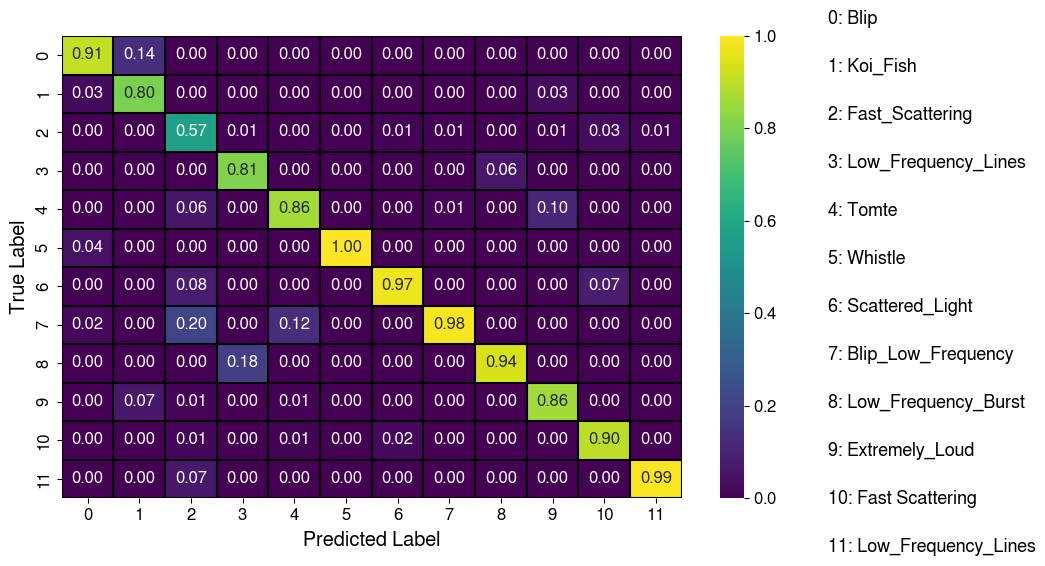}
    \caption{Confusion matrix illustrating glitch classification, employing Euclidean distance to determine the closest representative vectors from each class; the green and yellow cells signify a high degree of accuracy in predictions. On the right side, a list of correspondences to Gravity Spy classes is provided.}
    \label{fig:confusion_matrix}
\end{figure}

To classify an `unknown' glitch, we calculated the Euclidean distance between its vector and each class's representative vector in high-dimensional space (since, in principle, the unknown glitches lack t-SNE coordinates). The class with the closest distance was assigned to the glitch. Figure~\ref{fig:confusion_matrix} presents the classification results for the 20\% of data reserved as the test dataset, with predicted labels (horizontal axis) compared to true labels (vertical axis). 

As indicated by the green and yellow regions, most glitches were classified correctly. On the right, a list of corresponding Gravity Spy classes for each identified class is provided. The only class with lower classification accuracy was `Class 2', corresponding to the \textit{Fast Scattering} class, which contained the subdivision mentioned previously. The total accuracy for the normalized dataset, calculated as the sum of diagonal elements divided by the matrix total, is $88.33\%$. 

As mentioned before, the main idea here is to present the possibility of using the t-SNE output in different applications. In particular, this section shows that it is possible to have an automated classifier based on the two t-SNE coordinates, without the need for a training dataset. Although the results are interesting, for datasets with a large number of groups, the Euclidean distance may not be sufficient, since an unknown glitch could fall between two representative vectors. In this application, we could use the information to assign labels to the clusters, but alternative techniques - especially for large datasets - can be explored.

\section{t-SNE Applied to LLO Glitches During O3b without Prior Knowledge of Classes}
\label{sec:random_data}

\begin{figure}[ht!]
    \centering
    \begin{subfigure}[t]{0.48\textwidth}
        \centering
        \caption{t-SNE output for 2,000 glitches per month from November 2019 to March 2020}
        \includegraphics[width=\textwidth]{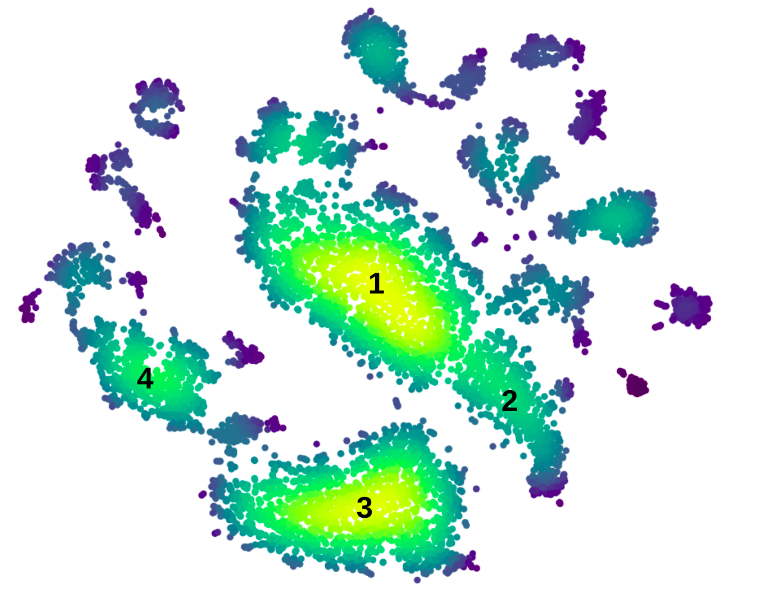}
        \label{fig:o3b_tsne_omicron_2000permonth}
    \end{subfigure}
    \hfill
    \begin{subfigure}[t]{0.48\textwidth}
        \centering
        \caption{t-SNE output for 10,000 transients randomly selected from the entire O3b}
        \includegraphics[width=\textwidth]{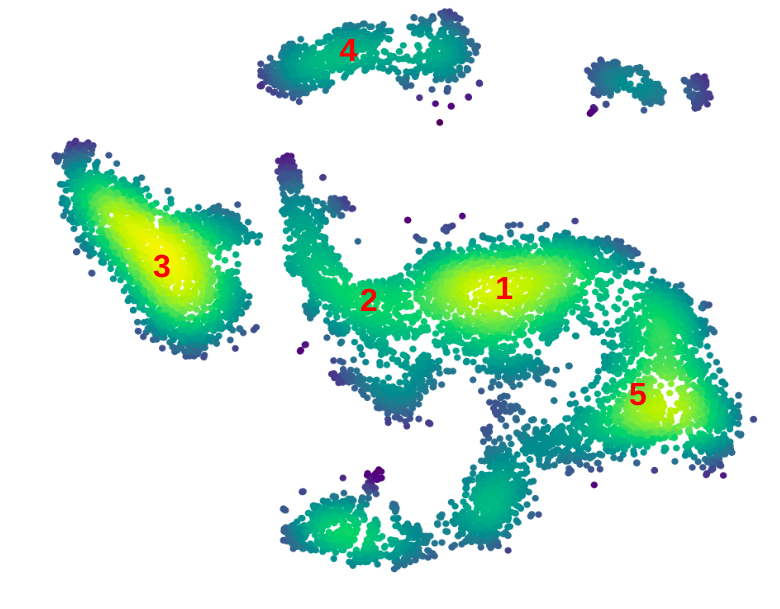}
        \label{fig:o3b_tsne_omicron_10000}
    \end{subfigure}
    \caption{Application of t-SNE to O3b LLO data without using class information. Yellow indicates regions with higher density.}
    \label{fig:}
\end{figure}

In this part of the analysis, we do not incorporate prior knowledge of glitch classifications. We randomly selected $2,000$ clustered Omicron transients per month during the O3b at LLO (from November 2019 to March 2020), filtering glitch frequencies between $10$ and $\SI{2048}{Hz}$, with a minimum SNR of $7.5$. We then applied t-SNE on these $10,000$ noise transients using unclustered Omicron information for each, normalized SNR, a window duration of $\SI{1}{second}$, and triggers for all $Q$-values. The resulting output is illustrated in Figure~\ref{fig:o3b_tsne_omicron_2000permonth}.

There are four main large groups and many other smaller ones. We analyzed the four main groups, which give the most common classes of glitches during O3b. Group 1, the largest, consists of the \textit{Fast Scattering} class with a frequency around $\SI{27}{Hz}$, which corresponds to the lower part of the red color in  Figure~\ref{fig:o3b_tsne_l1}. Group 2 also follows a similar distribution and is equivalent to the \textit{Fast Scattering} class. It exhibits the same characteristics in the upper part of the red color and it is in one of the new groups our method identified, with a higher frequency of approximately $\SI{39}{Hz}$. Group 3 is primarily composed of \textit{Tomte}, with \textit{Blip Low Frequency} on the right side. Finally, group 4 comprises glitches equivalent to \textit{Koi Fish} and \textit{Extremely Loud}. Overall, they follow the same characteristics as observed in Figure~\ref{fig:o3b_tsne_l1}. 

When considering glitches throughout the entire O3b, the \textit{Scattered Light} class was also among the most common glitch classes. However, our Omicron sample consisted of 2,000 observations per month and \textit{Scattered Light} class was concentrated in only some months. After the implementation of RC tracking~\cite{soni2020reducing} in January 2020, the number of glitches associated with this class decreased significantly, which explains the absence of a corresponding large cluster in the t-SNE output. Nevertheless, this class still appeared within the smaller groups in the upper right corner of the image. As mentioned earlier, each small group exhibits distinct characteristics, including varying durations, numbers of harmonics, and differences in the arches. 

If we select $10,000$ glitches randomly during all O3b, without filtering by month, this class, being one of the most common, becomes more prominent, as shown in Figure~\ref{fig:o3b_tsne_omicron_10000} in the region labeled $5$ (in red). This entire region and the lower left area contain \textit{Scattered Light} glitches, which are more concentrated than in the previous case. Each number in this new configuration corresponds to the groups in Figure~\ref{fig:o3b_tsne_omicron_2000permonth}; for example, number three in both cases includes more glitches classified as \textit{Tomte}. Note that the groups containing similar glitches appear similar in both images, aside from a possible rotation. Since we are selecting random data, the most common classes will naturally be predominant. Selecting an equal number per month enables a month-by-month analysis of glitch behavior, revealing when each group tends to dominate. Examples of glitch tracking across different time scales are provided in the following section.

\section{Applications: Exploring the Temporal Evolution of Glitches}
\label{sec:one_week_data}

\subsection{LLO - Analysis for One Week During O4a: Searching for Outliers}
\begin{figure}[ht!]
    \centering
    \includegraphics[width=0.99\textwidth]{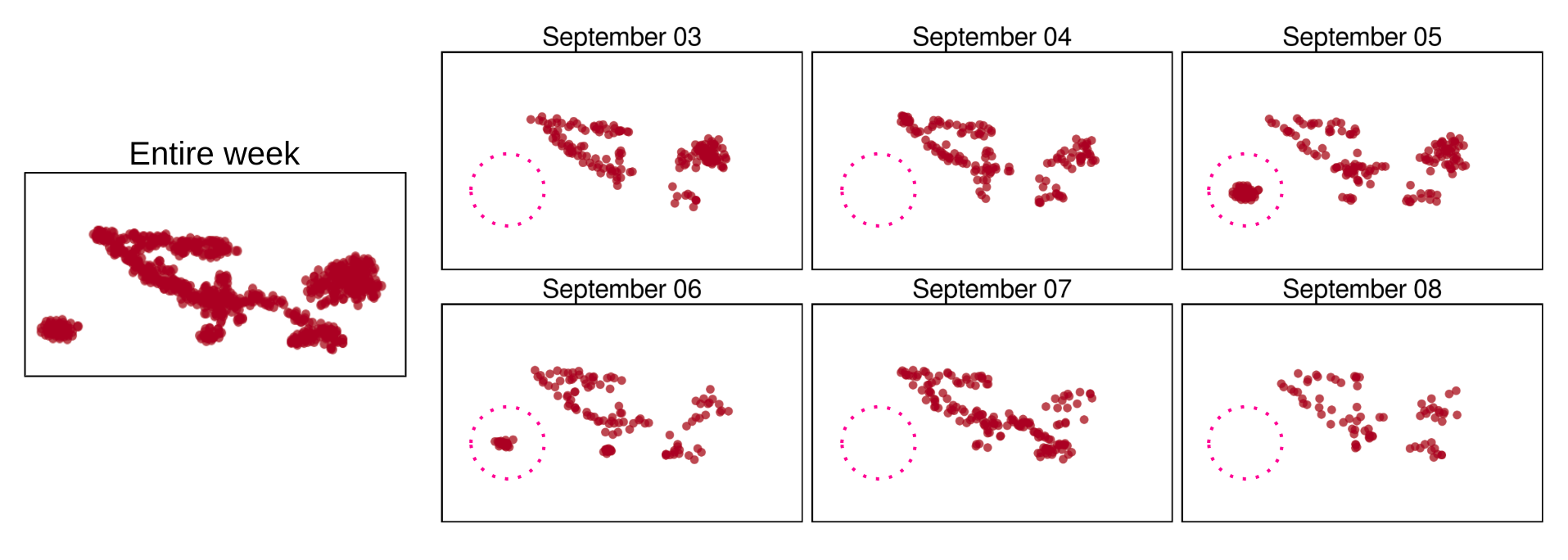}
    \caption{t-SNE visualization of data from September 3rd to 8th, 2024. On the left, the output for the entire week is presented, while on the right, the data is shown day by day, highlighting the formation and disappearance of a cluster (within the dashed circle).}
    \label{fig:daily_tsne}
\end{figure}

An advantage of this approach lies in its ability to analyze data with low latency and track changes over time. To verify this capability, we selected a week during O4a, specifically from September 3rd to September 8th, to investigate daily behavior without prior knowledge of the glitch classes. This allowed us to assess the natural occurrence and progression of glitches throughout the week. Figure~\ref{fig:daily_tsne} shows the application of t-SNE to the dataset, illustrating how the data evolves over the period from September 3rd to September 8th. 

The first image (on the left side) in \Fref{fig:daily_tsne} presents the output for the entire week, followed by the behavior observed on each day. Notably, a distinct cluster emerges on the left side of the plot after September 5th (highlighted by the dashed circle), shrinks on September 6th, and disappears by September 7th. This behavior suggests a temporal pattern linked to specific physical changes in the system. These glitches, observed at a frequency of $\SI{60}{Hz}$, were attributed to voltage bias changes in the electrostatic drive of the End Test Mass Y (ETMY) arm, as detailed in~\cite{67138}. This analysis demonstrates the potential to track temporal variations and associate them with environmental or instrumental conditions. The method can be applied on various timescales, including weekly, monthly, daily, or even hourly, depending on the specific objectives.

\subsection{LHO - Analysis for One Day During O4c: Hourly Evolution of Glitch Populations}
In this part of the analysis, we investigate how glitch populations - also without prior knowledge of classes - evolve over time by analyzing a single day of data from LHO during the O4c run. We selected all Omicron transients within a 24-hour period on February 21st, 2025. Since high-SNR glitches at Hanford occur at a relatively low rate, the only requirement applied was a minimum SNR of $6$, in order to retain a sufficient number of data points for running t-SNE and identifying clusters.

\begin{figure}[ht!]
    \centering
    \includegraphics[width=0.99\textwidth]{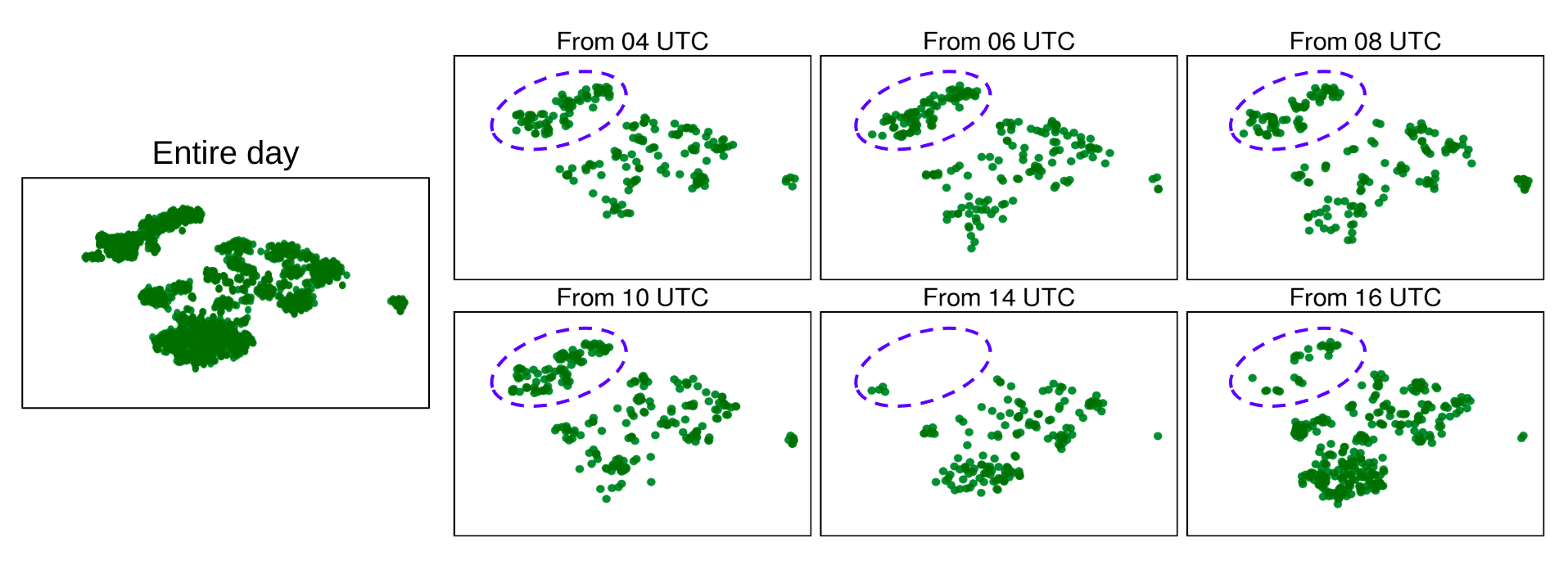}
    \caption{t-SNE visualization of data from February 21st, 2025. On the left, the output for the entire day at LHO is presented. On the right, the data is shown in two-hour segments (in UTC), highlighting the disappearance of a cluster after 14 UTC (within the dashed ellipse), which reappears around 17 UTC.}
    \label{fig:houly_tsne_lho}
\end{figure}

Unlike the previous application for LLO, here we explore how glitch morphology changes throughout the day by dividing the data into two-hour segments, starting from 04:00 UTC and continuing through 18:00 UTC. Figure~\ref{fig:houly_tsne_lho} shows the t-SNE output for the entire day on the left, while the remaining panels display each two-hour window separately. A particular cluster, highlighted by a dashed ellipse, appears prominently during the early hours (from 04 UTC onward) and becomes nearly absent between 14 UTC and 16 UTC. These glitches reappear around 17 UTC, as seen in the final panel covering the 16–18 UTC interval. According to~\cite{82959}, these glitches are correlated with control instabilities around 10.4 kHz.

The two observatory-based scenarios highlight the flexibility of this approach across different time scales and illustrate how certain glitch populations can temporarily disappear and reemerge over time. Such temporal dynamics may reflect changes in instrumental or environmental conditions and demonstrate the usefulness of the unsupervised method for tracking non-stationary glitch behavior.

\section{Discussions and Conclusions}
\label{sec:conclusions}

This paper presents a method for studying non-Gaussian noise transients (glitches) in gravitational wave detector data, focusing on the LIGO observatory through the examination of Omicron data. We employ clustering and classification techniques and investigate how variations in parameters influence the analysis. Since the primary dataset relies solely on unclustered Omicron information, the processes of time series acquisition, Fourier or $Q$-transforms for generating spectrograms, as well as the development of a training dataset, are not required. While supervised machine learning methods can yield improved classifications, establishing a robust training dataset poses significant challenges, often necessitating extensive time and frequent retraining to accommodate new classes.

The main idea is to explore the use of the t-SNE algorithm, a technique designed to visualize high-dimensional data in reduced dimensions by clustering data based on similarities, specifically applied to Omicron data to characterize and analyze glitches. In our initial application during O3b, we validated the t-SNE output against known glitch classes from Gravity Spy. We demonstrated that the most common classes were well grouped, except \textit{Koi Fish} and \textit{Extremely Loud} glitches. These two classes were identified as having similar morphologies, primarily differentiated by their SNR values, which significantly impact the results. When normalized SNR is not employed as input data, the \textit{Extremely Loud} class emerges as an outlier. Here, we chose to utilize normalized data to focus solely on glitch morphologies. If two glitches exhibit similar morphologies, they may originate from the same source, making it beneficial to group them together rather than separating them solely based on differences in SNR. However, the choice of approach should be guided by the primary objectives of the research.

We also conducted a study of the duration of the glitches, leading us to choose a window of $\SI{1}{s}$, which is large enough to retain information from classes such as \textit{Scattered Light} while being short enough to avoid mixing similar classes. This parameter is adjustable and, for instance, a longer window could be employed to study the repetition of scatter arches alongside low $Q$-values derived from the $Q$-transform, which allows for better temporal resolution. Our comparison of $Q$-values highlights how their variation affects the glitch morphology and the visualization in the t-SNE embedding. Higher $Q$-values (greater than 30), which provide poorer temporal resolution, demonstrate significant overlap among \textit{Tomte}, \textit{Fast Scattering}, and \textit{Blip Low Frequency}. In this context, distinct small clusters were identified for \textit{Low Frequency Burst}, which exhibited slight differences in frequency. This highlights a key feature of our methodology: the ability to select the range of $Q$-values based on the researcher’s interests. In our approach here, we utilized all $Q$-values, effectively summing all $Q$-planes from Omicron. This method captures triggers with lower resolution in time and frequency, but does so across all noise transients, ensuring that glitches with similar features are affected similarly.

Additionally, we demonstrated that it is possible to build a classifier using the t-SNE output, which is automated and does not require a training dataset. This classifier is based on the minimum Euclidean distance between an unknown glitch and the representative vectors created for each class (obtained from a combination of the Agglomerative Clustering and the Silhouette Score algorithms). The classifier was also applied to the same O3b data and identified two additional classes compared to Gravity Spy when the data were normalized, and one additional class when they were not. The main distinction arose between low-frequency glitches that differed in frequency, with one group centered around \SI{12}{\Hz}. It is important to emphasize that our goal was to illustrate the versatility of t-SNE visualizations, including their potential use in classification tasks. However, the main focus of this work is not to develop the best possible classifier, especially since we do not have a training dataset. Further investigations are planned, including exploring alternatives to Euclidean distance and testing different parameters. As mentioned, we are interested in the groups obtained from t-SNE.

We also explored the method applied to O3b without prior knowledge of glitch classifications, and we observed similar behaviors compared to the case with prior knowledge, particularly due to the morphologies of the most common glitches. One of the main objectives of this method is to present a means of tracking glitches over time and relating them to potential causes and sources. In this regard, we conducted a week-long analysis to identify outliers in the LLO data. By analyzing day by day, we observed a cluster of glitches shrinking, which were associated with instrumentation changes. This analysis can be performed on different time scales. For instance, we provide another application: an analysis using a two-hour time window with LHO data, where the identified group was associated with parametric instabilities. The tool can also be applied to Virgo and all other observatories where Omicron data is available, which is the only requirement for this method.

In conclusion, we present characteristics of an alternative approach to analyzing glitches, particularly in low-latency scenarios (relying on Omicron), by leveraging Omicron for rapid insights into glitch behavior. Although this method may not provide the highest resolution for classifying glitches, it facilitates a temporal understanding of glitch behavior, aiding in source identification. Furthermore, it eliminates the need to create a training dataset, and can be used for quick analysis during observing runs. This is intended to be improved for potential future applications, such as enabling the commissioning team to monitor glitch classes, and work on mitigation strategies. In future work, other techniques, such as UMAP, PCA, DBSCAN, or Deep Learning may be applied to further refine glitch classification, including the use of multiple data matrices to study glitches, testing with different durations, and varying $Q$-value ranges.

\ack{We would like to thank Jane Glanzer for the comments and suggestions. We express our gratitude to the members of the LIGO Detector Characterization Group and the LIGO-Virgo-KAGRA (LVK) collaboration for their valuable discussions. This work was supported by the National Science Foundation under grant numbers PHY-2110509 and PHY-2409740. The material is based upon work supported by NSF's LIGO Laboratory, a major facility fully funded by the National Science Foundation. Additionally, the work utilizes the LIGO computing clusters and data from the Advanced LIGO detectors; the authors are grateful for the computational resources provided by the LIGO Laboratory, supported by National Science Foundation Grants PHY-0757058 and PHY-0823459.}

\section*{References}

\bibliographystyle{iopart-num}
\providecommand{\newblock}{}

\end{document}